\documentclass[acmtog,screen,authorversion,nonacm]{acmart}
\acmSubmissionID{1423}

\usepackage{booktabs} 
\usepackage{multirow}
\usepackage{xcolor}
\usepackage{algorithm}
\usepackage{algpseudocode}
\usepackage{gensymb}

\acmJournal{TOG}

\copyrightyear{2026}
\acmYear{2026}
\setcopyright{cc}
\setcctype{by}
\acmConference[SA Conference Papers '26]{SIGGRAPH Asia 2026 Conference Papers}{December 01--04, 2026}{Kuala Lumpur, Malaysia}
\acmBooktitle{SIGGRAPH Asia 2026 Conference Papers (SA Conference Papers '26), December 01--04, 2026, Kuala Lumpur, Malaysia}
\acmDOI{10.1145/3829340.3842217}
\acmISBN{979-8-4007-2842-6/2026/12}

\begin{document}
\title{CuACD: A Fully GPU-Resident Approximate Convex Decomposition}

\author{Ruoxi Shi}
\affiliation{%
 \institution{UC San Diego}
 \streetaddress{9500 Gilman Dr}
 \city{La Jolla}
 \state{CA}
 \postcode{92093}
 \country{USA}}
\email{r8shi@ucsd.edu}
\author{Xinyue Wei}
\affiliation{%
 \institution{UC San Diego}
 \city{La Jolla}
 \country{USA}
}
\email{xiwei@ucsd.edu}
\author{Fanbo Xiang}
\affiliation{%
 \institution{Sudo GmbH}
 \city{La Jolla}
 \country{USA}
}
\email{fx@sudo.ai}
\author{Zexiang Xu}
\affiliation{%
 \institution{Sudo GmbH}
 \city{La Jolla}
 \country{USA}
}
\email{zexiangxu@sudo.ai}
\author{Hao Su}
\affiliation{%
 \institution{Sudo GmbH}
 \city{La Jolla}
 \country{USA}
}
\email{hao@sudo.ai}


\begin{abstract}
Approximate convex decomposition (ACD) converts triangle meshes
into small sets of convex parts and is a standard preprocessing
step for physics simulation and large-scale robot learning.  Modern
ACD methods produce high-quality decompositions through an
expensive search over candidate cutting planes, with per-mesh
runtime of tens of seconds that forces game pipelines into
offline asset processing.  Prior work has accelerated isolated stages, yet
the dominant costs---search, mesh cutting, and convex hull
construction---have remained on the CPU.

We adopt a two-layer concurrency view: a few hundred independent
tasks across the GPU's streaming multiprocessors, with each task
implemented as a warp---32 lanes executing in lockstep with
warp-level vote and shuffle intrinsics.  It fits the irregular
kernels of computational geometry that conventional thread-block
decomposition leaves underfilled, and dissolves the two inefficiencies
that kept ACD on the CPU: each kernel boundary ends in a tail of
stragglers that leaves most of the GPU idle, and variable-sized
phase outputs force the host CPU into the loop simply to size the
next launch.  A device-side heap allocator sizes inter-phase
buffers, allowing ACD's phases to fuse into warp-resident kernels
that never leave the GPU.  We present \emph{CuACD} (CUDA ACD), the
first fully GPU-resident ACD system, together with reusable GPU
components released as drop-in CUDA modules for search-based ACD
pipelines.  On the V-HACD benchmark, PartNet-Mobility, and an
Objaverse subset, CuACD achieves more than an order of magnitude of
speedup over prior baselines at matched or better quality.  The full
source code of the project is released at
\url{https://github.com/eliphatfs/cuacd}.
\end{abstract}

%
%


%
%


\begin{teaserfigure}
  \centering
  \includegraphics[width=\textwidth]{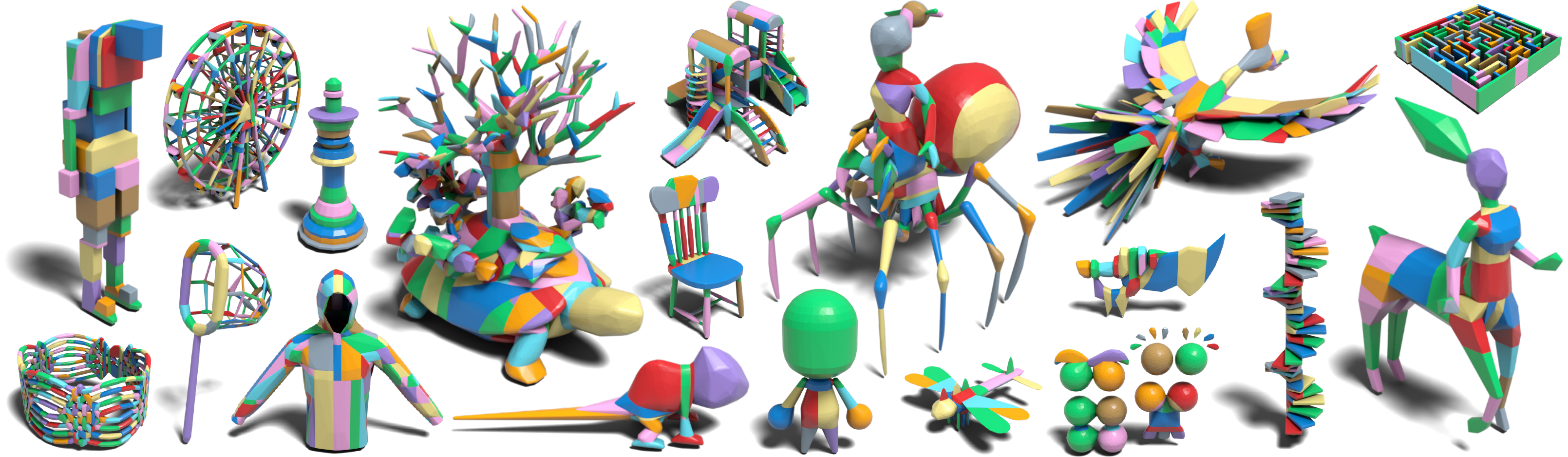}
  \caption{CuACD is a fully GPU-resident approximate convex
    decomposition method that delivers more than an order of
    magnitude of speedup over prior search-based ACD systems while
    faithfully preserving fine-grained geometric features.  The
    meshes shown here carry dense fine-grained structure and are
    decomposed at a tight concavity threshold $\tau\!=\!0.03$ (versus
    the $\tau\!=\!0.05$ default used in Table~\ref{tab:comparison}),
    yielding a higher per-mesh cost than the table reports; even on
    this harder regime, the entire gallery is produced in
    $17.71$\,s on a single NVIDIA RTX~4090 GPU. The source code is publicly available at \url{https://github.com/eliphatfs/cuacd}.}
  \label{fig:teaser}
\end{teaserfigure}

\maketitle


\section{Introduction}
\label{sec:intro}

Approximate convex decomposition (ACD) is a core primitive of
physical simulation and interactive 3D graphics. Modern physics
engines rely on convex colliders because intersection and contact
queries against convex shapes admit fast, near-constant-time
algorithms~\cite{Gilbert1988GJK,Mirtich1998VClip} that have no
counterpart on arbitrary triangle meshes;
consequently, almost every character, prop, and environment asset in
a modern video game or robotic simulator must be decomposed into a
small set of convex parts before it can be used at
runtime~\cite{Coumans2015Bullet,Tao2024ManiSkill3}.

Exact convex decomposition is NP-hard~\cite{Chazelle1984ConvexPartition}
and, in practice, produces partitions with orders of magnitude more
pieces than downstream simulators require~\cite{Lien2007ACDPolyhedra}. A
decade of research has therefore converged on approximate methods:
HACD~\cite{Mamou2009HACD}, V-HACD~\cite{Mamou2016VHACD},
CoACD~\cite{Wei2022CoACD}, NavACD~\cite{Lands2024NavACD}, and the
recent VisACD~\cite{Fokin2026VisACD}. These works have progressively
improved decomposition quality; while HACD is clustering-based, the
majority of methods since V-HACD share a common computational
backbone: an expensive search over candidate cutting planes evaluated
by an equally expensive \emph{concavity} objective---a scalar
measuring how far a piece deviates from convex.

The speed of that search has become a real bottleneck in modern 3D
content pipelines. Large-scale robot-learning efforts routinely
preprocess articulated-object datasets such as
PartNet-Mobility~\cite{Xiang2020PartNet} into convex colliders before
a simulator can consume them~\cite{Tao2024ManiSkill3}; with single-mesh
decomposition times of tens of seconds, even
medium-sized datasets can occupy a CPU cluster for days, and recent
large-scale 3D~AI pipelines repeatedly identify ACD as a dominant
preprocessing cost~\cite{Tao2024ManiSkill3,Fokin2026VisACD}. In
large-scale game development, where artists iterate on hundreds of
mesh assets per sprint, convex decomposition is typically delegated to
offline asset processing rather than integrated into interactive authoring,
slowing iteration and encouraging workarounds such as hand-authored
proxy geometry.

ACD can be formulated as a search problem: given an input mesh, find a
sequence of planar cuts that minimizes a concavity objective while
respecting a budget on the number of parts. Evaluating that objective
at each candidate cut requires building the convex hulls of the two
resulting pieces and measuring their deviation from the input surface.
Across a single decomposition this inner loop executes hundreds of
thousands of times, making cutting and convex hull construction the
clear computational hot spots. VisACD takes a first step toward GPU
acceleration by offloading ray-based visibility queries onto CUDA and
OptiX, achieving roughly a $2\times$ speedup over
CoACD~\cite{Fokin2026VisACD}; intersection tests and convex hull
construction, however, remain on the CPU, leaving the dominant cost of
an ACD pipeline untouched. We instead present \emph{CuACD} (CUDA
ACD), a \emph{fully} GPU-resident ACD algorithm: once the input
mesh is uploaded, the worklist, lookahead tree, cutting, hull, and
concavity all live in device memory, and the host issues no
intervening synchronization until the decomposition is read back.
This yields orders of magnitude of speedup over CoACD.
Porting ACD to the GPU is far from mechanical.  The irregular,
data-dependent control flow of convex hull construction, polygon
boundary tracing, ear-clip triangulation, and recursive tree search
is a poor fit for the GPU's SIMT execution model---single
instruction, multiple threads, in which $32$ hardware lanes share
one instruction pointer and advance together each
cycle~\cite{Lindholm2008Tesla}.  The natural CUDA response is to
split the pipeline phase by phase into a sequence of small,
homogeneous kernel launches, but this trades one problem for two.
First, each kernel ends with a tail of stragglers: as the work
distribution shrinks and reshapes between phases, only a few lanes
have remaining work, and most of the GPU sits idle until the next
launch~\cite{Aila2009RayTraversal}.  Second, each phase's
variable-sized output is known only after the producing kernel
finishes, so the host CPU must be involved between phases simply
to read the output size and allocate the next kernel's input.  Both
costs scale with the number of kernel boundaries, and together they
break the savings that a GPU algorithm was meant to deliver.

\begin{figure}[t]
  \centering
  \includegraphics[width=\linewidth]{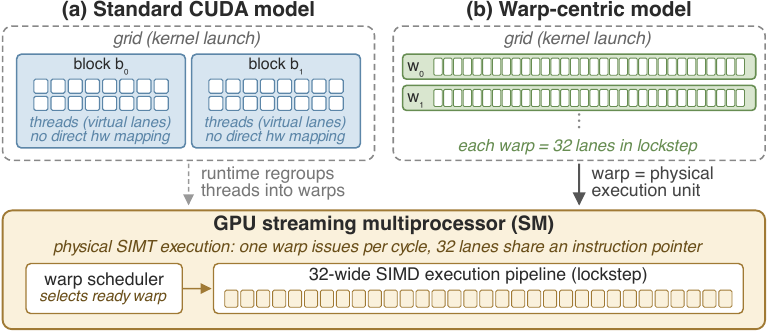}
  \caption{CUDA exposes a three-level grid/block/thread hierarchy in which threads and blocks are virtual abstractions implicitly regrouped by the runtime, while only the warp has a hardware counterpart on the streaming multiprocessor (left). The warp-centric view (right) elevates the warp itself to the unit of algorithm design: each warp is a 32-lane SIMD processor mapped one-to-one onto the SM's physical execution pipeline.}
  \label{fig:cuda-vs-warp}
  \vspace{-2em}
\end{figure}

A cleaner fit is to treat the warp, rather than the thread or thread
block, as the unit in which GPU algorithms are expressed
(Figure~\ref{fig:cuda-vs-warp}).  CUDA's programming model exposes a
three-level hierarchy of grid, block, and thread.
People familiar with the CPU may assume that we have hundreds of thousands of independent thread resources on a modern GPU.
However, only the \emph{warp}---a fixed group of $32$ lanes sharing one
instruction pointer---has a hardware counterpart on the streaming
multiprocessor; threads and blocks are virtual abstractions whose
mapping to physical lanes is implicit.

Adopting the warp as the unit of algorithm design yields a
two-layer concurrency picture.  On the coarse level, an algorithm exposes a few
hundred independent tasks, each pinned to one warp and dispatched
across multiprocessors.  On the fine level, each warp is a $32$-lane SIMD
processor whose lanes cooperate through warp-level
intrinsics.
The fine cooperation allows much greater flexibility of processing on most programs, even for workloads with sequential dependencies --
this is the same idiom by which workloads long held to
be inherently sequential---JSON parsing~\cite{Langdale2019simdjson}
and multi-pattern regex matching~\cite{Wang2019Hyperscan}---now run
on SIMD hardware at gigabytes per second.  The parallelism an
algorithm must expose drops accordingly.  Filling a modern GPU at
thread granularity needs ${\sim}10^{4}$ independent threads, but at
warp granularity the same machine schedules only ${\sim}10^{2}$
resident warps---a budget the $10^{2}$--$10^{3}$ active parts of an
ACD worklist meet with room to spare.  The view was introduced for
irregular graph traversal by \citet{Hong2011WarpCentric}; we adapt
it here to code up a computational-geometry pipeline stages into warp-resident kernels,
paired with a device-side heap allocator that sizes and places the
variable-sized inter-phase buffers on the device.  We develop both
in Sec.\@~\ref{sec:warp}.

In summary, this paper makes the following contributions:
\begin{itemize}
  \item A warp-centric design for the irregular,
    data-dependent kernels of computational geometry: it adapts the
    warp-unit view of \citet{Hong2011WarpCentric} from graph
    traversal, and lowers the concurrency that a SIMT
    mapping must expose from ${\sim}10^{4}$ independent threads
    to ${\sim}10^{2}$ resident warps treated as SIMD processing units,
    and is paired with a device-side heap allocator so that an entire
    pipeline can run without leaving the GPU.
  \item The first fully GPU-resident ACD algorithm and system.%
  \item A suite of fast, reusable GPU components for computational
    geometry---a device-side heap allocator, warp-level
    quick-sort, mesh--plane clipping, convex hull construction,
    connected-component analysis, and Hausdorff distance---released
    as standalone CUDA modules that drop into existing search-based
    ACD pipelines beyond CuACD; the full source code of the project is
    released at \url{https://github.com/eliphatfs/cuacd}.%
  \item Orders of magnitude of speedup over CoACD, while
    also achieving better decomposition quality.
\end{itemize}

\section{Related Work}
\label{sec:related}

\paragraph{Approximate convex decomposition.}
ACD was introduced by \citet{Lien2004ACD} for polygons and extended to
3D polyhedra by \citet{Lien2007ACDPolyhedra}.
HACD~\cite{Mamou2009HACD} operates directly on the triangle mesh,
building a dual graph over triangle clusters and iteratively merging
adjacent clusters so as to minimize the concavity of each cluster's
convex hull.  \citet{Attene2008Hierarchical} offer a bottom-up
counterpart, computing hierarchical convex approximations by
tetrahedralizing the mesh.
V-HACD~\cite{Mamou2016VHACD} replaces triangle-level clustering with a voxelized recursive bisection that repeatedly picks
axis-aligned splitting planes minimizing a volumetric concavity
objective, trading some accuracy for substantial gains in robustness
and speed, and in the process becoming the de-facto default across
game and simulator pipelines; \citet{Muller2013Fracture} adapt this
volumetric formulation to a real-time GPU implementation for dynamic
fracture, and \citet{Thul2018ACDAnimation} extend ACD to animated
meshes via temporal coherence, while also introducing concave-edge plane
sampling for ACD. CoACD~\cite{Wei2022CoACD} introduces a ``collision-aware'' concavity built from the bidirectional Hausdorff
distance so that
contact-relevant surfaces dominate the score, and drives the search with Monte Carlo tree
search, markedly lifting decomposition quality at the cost of a longer
runtime. NavACD~\cite{Lands2024NavACD} specializes ACD to navigation
use cases: it first infers a navigable free-space region and then
constrains the decomposition to avoid intruding into it, producing
fewer and larger hulls well suited to game AI and robot locomotion.
VisACD~\cite{Fokin2026VisACD} adopts a visibility-based concavity
metric---rotation-equivariant in the sense that rotating the input
rotates the score commutatively---evaluated with OptiX ray queries
on the GPU, reporting roughly ${\sim}2\times$ end-to-end speedup over CoACD;
however, cutting, hull construction, and the outer search all
remain on the CPU. Our system belongs to the same search-based
lineage but is the first to be \emph{fully} GPU-resident---cutting,
hull construction, concavity evaluation, and tree search all execute
on device---yielding more than an order of magnitude of throughput
improvement while also producing decompositions of higher
quality.

\paragraph{Geometry processing on the GPU.}
A line of work starting with CUDARaster~\cite{Laine2011CUDARaster}
and parallel bounding volume hierarchy
construction~\cite{Lauterbach2009LBVH,Karras2012BVH} established that
core geometric primitives---rasterization, spatial indexing,
hierarchy building, and traversal---can be mapped efficiently onto the
GPU's SIMT execution model, typically by recasting irregular
computations as parallel sorts on spatial keys, parallel construction of
bounding-volume hierarchies (BVHs), or tile-parallel reductions. More recent efforts have pushed this agenda
into heavier, globally data-dependent operations: NVIDIA's lock-free
interactive decimation of large meshes~\cite{Gautron2023Decimation},
the PaMO pipeline for parallel intersection-free remeshing and mesh
simplification~\cite{Oh2025PaMO}, and the aforementioned
VisACD~\cite{Fokin2026VisACD}.  \citet{Mahmoud2021RXMesh} offer a
general GPU mesh data structure that accelerates neighbor queries and
dynamic updates; CuACD's kernels instead work on plain
struct-of-arrays buffers so the components stay drop-in.  Convex hull
construction has also received dedicated GPU treatments~\cite{Tang2012GPUHull,Gao2013gHull},
but these algorithms extract parallelism \emph{within} a single large
point cloud (one hull per kernel launch), whereas ACD's inner loop
instead demands thousands of tiny hulls of tens to a few thousand
points each, built concurrently---a regime in which per-hull launch
overhead and block-level coordination dominate.  Our work continues
this trajectory and extends it to the ACD pipeline as a whole,
including convex hull construction and lookahead tree
search---computations previously considered a poor match for SIMT
execution.

\paragraph{Learning-based convex decomposition.}
A complementary line of work uses neural networks to propose cutting
planes~\cite{Luo2025RLACD}, to represent shapes as differentiable
unions of convex primitives~\cite{Deng2020CvxNet}, or to cluster
learned feature fields into convex
parts~\cite{Yang2026FeatureFieldACD}.  These methods accelerate the
\emph{heuristic} via GPU-resident neural inference; we instead
accelerate the \emph{classical search itself} with lightweight
hand-designed kernels, and the two strategies are complementary.

\section{A Warp-Centric View of Geometry Processing}
\label{sec:warp}

The view that the warp should be the unit in which GPU algorithms
are expressed is not new: \citet{Hong2011WarpCentric} introduced a
virtual warp-centric method for irregular \emph{graph} algorithms,
where intra-warp load imbalance from skewed vertex degrees was the
dominant pathology, and \citet{Zou2024MetaMeshing} apply the
term to cache-aware data layout for lattice triangulation.
We adopt the warp as the unit of algorithm design in the same spirit,
but transfer the view to computational geometry and apply it to CuACD.
Once the warp is taken as the unit, the GPU exposes two layers of
concurrency (Sec.\@~\ref{sec:intro}): a coarse layer, across which
independent tasks are dispatched to streaming multiprocessors, and a
fine layer, in which each task runs as a $32$-lane warp in lockstep
SIMD.  Two facts about ACD then determine how the algorithm fills these
two layers and where the design effort falls.  \emph{(i)}~The active
worklist holds $10^{2}$--$10^{3}$ independent parts, so the coarse layer
is saturated.  \emph{(ii)}~Evaluating one candidate cut is irregular,
variable-size, data-dependent work, so the fine layer is the hard part.
Sec.\@~\ref{sec:warprecipe} develops the fine layer; Sec.\@~\ref{sec:heap}
then supplies the device-resident memory that lets the two layers fuse
without leaving the GPU.%

\paragraph{Concurrency is the central resource.}
GPU throughput is governed by Little's law: sustained arithmetic rate
equals in-flight instructions divided by their average latency.
The unit in which we count that concurrency---thread, warp, or
block---determines which algorithms we consider GPU-amenable:
as Sec.\@~\ref{sec:intro} noted, filling a modern GPU at thread
granularity needs ${\sim}10^{4}$ independent threads, but at warp
granularity the same machine schedules only ${\sim}10^{2}$ resident
warps---a budget the $10^{2}$--$10^{3}$ active parts of an ACD
worklist meet with room to spare.  CuACD therefore counts concurrency
at the warp.
The coarse layer is the easy half: search-based algorithms expose it
trivially, and in CuACD the active worklist and the per-part clipping
candidate pool together saturate it. We detail the fine layer in the following subsection.

\subsection{Fine Layer: Evaluating One Candidate as a Warp}
\label{sec:warprecipe}

The fine layer---coding one candidate's evaluation as a $32$-wide
warp---is where the warp-centric design earns its keep, because that
work is irregular, variable-size, and data-dependent.  We decompose the
body of one candidate evaluation into phases drawn from the following
four primitives, in the listed order of preference; each inherits a
structural parallel-programming pattern long familiar from the CPU
literature, and each has a tight warp-level implementation illustrated
in Appendix~I of the supplemental.%
\begin{enumerate}
  \item \emph{Traversing an array.} The 32 lanes traverse the array
    cooperatively at stride 32.  We support three modes of operation:
    Map, Reduction, and Filtering. See Appendix~I for detailed instructions.
  \item \emph{Sorting and set operations.}
    These reduce to the warp-cooperative quick-sort of
    Sec.\@~\ref{sec:sort}, whose three-way partition is itself a
    filter (case~1).  Set construction or deduplication extends sort with a filter on
    adjacent duplicates; membership appends a binary search.
  \item \emph{Fine-grained symmetric work.} Operations that act in
    parallel on a pair of related entities---both directions of a
    doubly linked list, or both endpoints of an edge---partition the
    warp into 16 two-lane groups with high SIMD efficiency.  CuACD
    uses this in the convex-hull merge of Sec.\@~\ref{sec:hull}.
  \item \emph{Divide and conquer.} A 32-way split assigns one branch
    per lane and recombines the results afterwards; we apply this to
    recursive sub-hull construction (Sec.\@~\ref{sec:hull}).  It is
    considered last because the data-dependent control flow inside
    each branch is not guaranteed to retain full warp utility.
\end{enumerate}
Every primitive is instantiated by a concrete CuACD stage: the
mesh-plane clipper is a filter over crossing edges followed by a
sort-and-dedup (primitives~1--2); the convex-hull build is the
divide-and-conquer primitive~(4), with its merge using the
symmetric-work primitive~(3); the concavity surrogate is a warp
reduction (primitive~1).  The structure of Sec.\@~\ref{sec:method} is
thus the structure of this list.%

\paragraph{Meta-ops.} For metadata tracking operations that do not scale with the input size, we simply do them on thread 0 of a warp with negligible performance penalty.

The existing literature~\cite{Blelloch1989Scans,Cole1989Skeletons,Dean2008MapReduce}
suggests that these aspects already cover most patterns in a practical algorithm.

Our implementation has two advantages over existing libraries like CUB~\cite{CUB}.
Firstly, we do not require static-sized input for any of the primitives, which is fine for use in a thread block, but for warp-centric algorithms, organization becomes impractical.
Secondly, we do not require temporary scratch in shared memory, which does not scale well with the number of primitive phases in the algorithm or the size of the input (see Appendix~Table~3).

\paragraph{Primitive~(2) illustrated: warp-cooperative quick-sort.}
\label{sec:sort}
GPU sorting usually favors bitonic or radix sort~\cite{yildiz2013parallelization},
less often merge sort~\cite{CUB}; \citet{cederman2010gpu} deemed
quick-sort impractical on GPUs and offloaded its work assignment to the
CPU.  Yet quick-sort aligns naturally with our warp-centric model: its
partition step is exactly the warp-cooperative filter of
primitive~(1), so it inherits the warp model's efficiency without the
shared-memory and static-size constraints that make bitonic and radix
sorts awkward for the small, variable-length segments ACD
produces.
We naturally benefit from its cache friendliness.

We maintain a per-warp stack of segments awaiting partition, initialized with the full input.
For each segment to process: if the segment has 32 or fewer elements, we apply a bitonic network to sort them directly.
Otherwise, we pick a pivot by taking a systematic sample of size 32 from the array to partition, and using the same bitonic network to find the median across the warp as the pivot for partitioning. After that, the warp cooperatively scans and partitions the segment using warp-vote intrinsics (Appendix~I).

\subsection{Device-Resident Fusion: The Heap Allocator}
\label{sec:heap}

\begin{figure}[t]
  \centering
  \includegraphics[width=\linewidth]{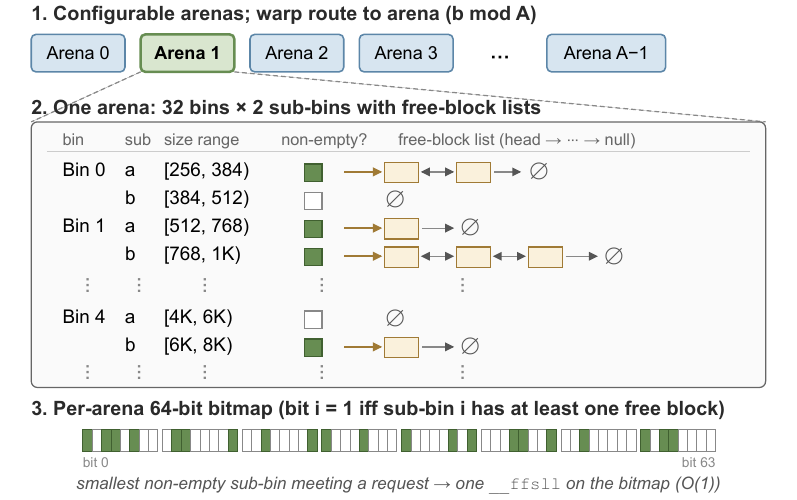}
  \caption{Device-side heap allocator.}
  \label{fig:heap-allocator}
\end{figure}

The coarse and fine layers together make each candidate evaluable as a
single warp, but for the warp-granularity saturation above to actually
pay off, the per-warp kernels must fuse into one device-resident
pipeline without a host round-trip between phases---the very cost that,
as Sec.\@~\ref{sec:intro} noted, breaks a naive phase-by-phase CUDA
port.  The piece that removes the last host dependency is a
device-side heap allocator, because many of CuACD's intermediate
buffers have sizes known only after the producing kernel runs.
Examples include vertex arrays from
plane cuts, edge and face arrays for hull construction, and BVH
scratch for Hausdorff queries.  CUDA's built-in device-side \texttt{malloc}/\texttt{free} avoids
the host synchronization but serializes through a single global
structure.  At CuACD's allocation rate, this serialization becomes
the dominant cost (quantified in Sec.\@~\ref{sec:components}).
Existing GPU allocators such as
ScatterAlloc~\cite{Steinberger2012ScatterAlloc} target small-object
fast paths and do not coalesce freed blocks.  CuACD's variable-size
mesh buffers therefore fragment their heap quickly.  The remedy is
the coalescing-binned machinery---classical CPU allocator design,
inherited from Hoard~\cite{Berger2000Hoard},
dlmalloc~\cite{Lea2000dlmalloc}, and Knuth's boundary-tag
scheme~\cite{Knuth1997TAOCP1}.  Our contribution is the adaptation
of that machinery to a warp-scheduled SIMT execution model, in which
the relevant concurrency unit is a warp of co-issued lanes, not a
thread.

\paragraph{Structure.}
The heap splits into $A$ independent arenas; a request from warp~$b$ routes to arena $b\bmod A$, with $A{=}64$ sized to keep
the critical path contention-free at CuACD's occupancies
(Sec.\@~\ref{sec:components}).  Each arena holds $32$ power-of-two
bins split at the midpoint into $64$ sub-bins, each owning a
linked free list (Figure~\ref{fig:heap-allocator}); a
per-arena $64$-bit bitmap flags non-empty lists, so a fitting
sub-bin is found in one bitwise operation and on free boundary
tags coalesce with neighbors in $O(1)$.
This is the mechanism that lets the entire decomposition stay
device-resident: the variable-size vertex arrays a plane cut emits, the
edge and face arrays hull construction needs, and the BVH scratch a
Hausdorff query needs are all allocated and freed on-device, so the
host issues no synchronization between the initial upload and the final
read-back.
More details can be found in Appendix~E of the supplemental.
\section{The CuACD Algorithm}
\label{sec:method}

\subsection{The CuACD Pipeline}
\label{sec:pipeline}

\begin{figure}[t]
  \centering
  \includegraphics[width=0.98\linewidth]{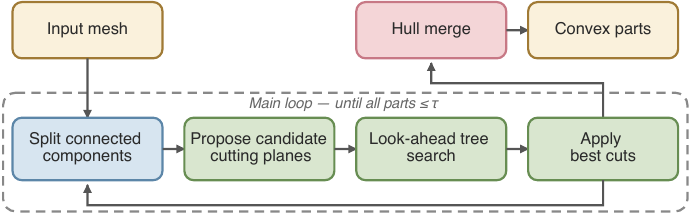}
  \caption{Overview of the CuACD decomposition pipeline.  Input mesh
    is decomposed by an outer loop that splits
    multi-component parts, evaluates concavity, and selects the best
    cut for any part exceeding the threshold $\tau$ via a bounded
    look-ahead tree search; a
    greedy hull-merge pass reduces the final part count after the loop.}
  \label{fig:pipeline}
  \vspace{-1.4em}
\end{figure}

CuACD follows the overall structure of CoACD's search-based
decomposition~\cite{Wei2022CoACD}, but redesigns every stage so that
the entire pipeline runs on the GPU.
Figure~\ref{fig:pipeline} summarizes the pipeline.
The algorithm maintains a \emph{worklist} of parts that still require
further decomposition, seeded with the input mesh as a single
element.  At each iteration, in parallel, we select every part whose concavity exceeds~$\tau$.  For each
selected part we choose a cutting plane via a tree search, split the
part along that plane, and push the resulting sub-parts back onto the
worklist if they remain above the threshold.  The loop terminates
when no part remains in the worklist.  Then, a hull-merge
post-processing pass, in the spirit of CoACD's
merge step, greedily combines pairs of output parts whose joint
convex hull remains below the concavity threshold, reducing the
final part count without regressing quality; we describe the
GPU-resident version of this pass in Sec.\@~\ref{sec:hullmerge}.

Choosing a cutting plane has two phases: a \emph{proposal} phase that enumerates a pool of candidate planes, and an \emph{evaluation} phase that scores each candidate by a parallel look-ahead tree search and returns the lowest-cost plane.

\paragraph{Plane proposal.}
We combine two candidate families. \emph{Axis-aligned planes} sample each principal axis of the part's bounding box at $W_{\mathrm{top}}$ evenly spaced positions, as in CoACD. \emph{Concave-edge planes} target contact-relevant features: following \citet{Thul2018ACDAnimation}, we sample up to $N_{\mathrm{ce}}$ concave edges with dihedral angle $\ge 200\degree$ and emit four candidate planes per edge---two aligned to the dihedral bisector and shifted by $\epsilon_{conc}$ in both normal directions, two directly aligned to the incident faces.  The two families form a pool of $W_{\mathrm{top}} + 4\,N_{\mathrm{ce}}$ at the tree-search root.%
\paragraph{Concavity metric.}
We adopt CoACD's concavity metric unchanged.  For a part~$P$ with
convex hull~$H(P)$, the concavity is
\begin{equation}
  \mathrm{cost}(P) \;=\; \max\bigl(k\,R_v(P),\; d_H(P,H(P))\bigr),
  \label{eq:cost}
\end{equation}
where $R_v$ is the radius of a sphere matching the residual volume
between hull and mesh, $d_H$ is the bidirectional Hausdorff distance,
and $k$ is a calibration coefficient~\cite{Wei2022CoACD}.  
The two are complementary: $kR_v$ captures missing hull volume, while $d_H$ captures the worst surface deviation in either direction.
Similar to CoACD, during the tree search we use the cheaper $kR_v$ surrogate; the full metric is evaluated only at the termination test.

\paragraph{Parallel look-ahead evaluation.}
Each candidate plane is evaluated via a bounded-depth tree search that exposes two levels of GPU parallelism: across parts being cut concurrently, and across candidates within each part. To evaluate a candidate, we apply the cut, hull and score the two resulting sub-parts by the surrogate cost $R_v$, then recursively expand the worse sub-part to depth~$D$ using a reduced axis-aligned branching factor $W_{\mathrm{sub}}$. The branch whose path-average worst-sub-part cost is lowest defines the selected cut.%

\subsection{Mesh--Plane Clipper}
\label{sec:clip}

\begin{algorithm}[!b]
\caption{Mesh--plane clipping.
  Each step is tagged by its warp-centric category from
  Sec.\@~\ref{sec:warprecipe}.}
\label{alg:clip}
\small
\begin{algorithmic}[1]
  \Require{Vertex set $V$, Triangle set $T$, Cutting plane $\pi$.}
  \Ensure{Output meshes $M^{\pm}$.}
  \State \textcolor{blue}{// Intersect with cutting plane.}
  \State $\mathrm{signs}[v]\!\gets\!\mathrm{sign}(\pi\!\cdot\!v), \forall v \in V$
         \Comment{(1) Map}
  \State $X\!\gets\!\{(\min,\max)$ keys of edges of $T$ crossing $\pi\}$
         \Comment{(1) Filter}
  \State $X\!\gets\!\Call{Sort}{X}$
         \Comment{(2) Sort (for set building)}
  \State Add intersection vertex per unique key of $X$
         \Comment{(2) Dedup; (1) Filter}
  \State \textcolor{blue}{// Build positive and negative parts without cap.}
  \State Split each $t\!\in\!T$; append to $T^{+},T^{-}$
         \Comment{(1) Filter}
  \State \textcolor{blue}{// Find boundary edges for cap ear-clipping.}
  \State $D\!\gets\!\{(\min,\max)$ keys of edges of the smaller of $T^{\pm}\}$
         \Comment{(1) Map}
  \State $D\!\gets\!\Call{Sort}{D}$
         \Comment{(2) Sort (for set building)}
  \State $B\!\gets\!\{D[i]\!:\!D[i]\text{ unpaired with sort neighbor}\}$
         \Comment{(2) Dedup}
  \State \textcolor{blue}{// Build the intersection caps.}
  \State Chain $B$ into closed loops $L_{1},\dots,L_{k}$
         \Comment{(1) Reduce}
  \State Classify each $L_{i}$ into outer/inner loop by 2D ray-casting.
         \Comment{(1) Map}
  \State Bridge holes by ear-clipping into cap $C$
         \Comment{(1) Reduce}
  \State \textcolor{blue}{// Write back.}
  \State Heap-allocate output buffers
         \Comment{Meta}
  \State Compact vertices, remap indices, write $M^{+},M^{-}$
         \Comment{(1) Map}
\end{algorithmic}
\end{algorithm}

For each branching node and each candidate plane, the look-ahead evaluator performs one mesh–plane cut. This produces two watertight sub-meshes, with the cut section re-triangulated to close the boundary. Algorithm~\ref{alg:clip} gives the
pipeline; scratch space is allocated from the heap of Sec.\@~\ref{sec:heap}. Allocation is seen as meta operations executed on lane 0 as the time do not scale with input size.

The algorithm follows a standard scheme: compute each vertex's sign
against the cutting plane, then split the mesh into $T^{+}$ and $T^{-}$
by that sign.  Step~7 produces the two triangle sets but does not mark
which edges lie on the cut, so the boundary must be recovered.  We use
the manifold property---an interior edge is shared by exactly two
triangles: building the sorted edge-key multiset of the smaller part,
any key appearing once is a boundary edge, and steps~9--11 construct
that multiset and isolate the unpaired keys into the boundary set~$B$
that the cap-loop chaining consumes.

The cap itself uses standard ear-clipping.  We classify inner and outer
loops on the cutting plane by ray-casting from the first vertex and
counting outer-loop crossings; all primitives here are linear
traversals parallelizable by item~1 of~\ref{sec:warprecipe}. Appendix~F walks the cap step by step.

\subsection{Convex Hull and Concavity Evaluation}
\label{sec:hull}

Every concavity computation requires hulling the part. A single decomposition invokes the hull kernel hundreds of thousands of times. Each call operates on anywhere from a few dozen to tens of thousands of points. We ship two exact variants sharing a
divide-and-conquer core: a plain \emph{D\&C} path and a
\emph{hybrid} path that prepends a direction-extreme prefilter.

\begin{table*}[t]
\small
\caption{Concavity, number of parts, and decomposition time across
  three benchmarks, averaged over the per-mesh results.  All
  baselines are operated at the threshold sweep that targets CoACD's
  default mean concavity ($0.05$); residual concavity differences
  reflect the granularity of each system's threshold knob rather than
  a quality gap.  Bold marks the best entry per column; lower is
  better on all three axes.}
\label{tab:comparison}
\centering
\begin{tabular}{lccccccccc}
  \toprule
   & \multicolumn{3}{c}{V-HACD} & \multicolumn{3}{c}{PartNet-Mobility} & \multicolumn{3}{c}{Objaverse subset} \\
  \cmidrule(lr){2-4} \cmidrule(lr){5-7} \cmidrule(lr){8-10}
  \textbf{Method}
    & \textbf{Concavity} & \textbf{\#Parts} & \textbf{Time (s)}
    & \textbf{Concavity} & \textbf{\#Parts} & \textbf{Time (s)}
    & \textbf{Concavity} & \textbf{\#Parts} & \textbf{Time (s)} \\
  \midrule
  CoACD~\cite{Wei2022CoACD}     & $0.0495$ & $40.5$ & $18.03$ & $0.0465$ & $21.1$ & $12.82$ & $0.0540$ & $59.0$ & $25.93$ \\
  NavACD~\cite{Lands2024NavACD} & $0.0529$ & $79.1$ & $12.13$ & $0.0614$ & $25.6$ & $2.96$  & $0.0950$ & $79.5$ & $40.25$ \\
  VisACD~\cite{Fokin2026VisACD} & $0.0604$ & $37.5$ & $9.60$  & $0.0511$ & $22.8$ & $7.42$  & $0.0702$ & $59.9$ & $15.92$ \\
  Ours                          & $\mathbf{0.0488}$ & $\mathbf{33.6}$ & $\mathbf{0.23}$  & $\mathbf{0.0458}$ & $\mathbf{21.0}$ & $\mathbf{0.16}$  & $\mathbf{0.0496}$ & $\mathbf{48.9}$ & $\mathbf{0.25}$ \\
  \bottomrule
\end{tabular}
\vspace{-0.5em}
\end{table*}

\paragraph{Divide-and-conquer core.}
The core is a warp-cooperative GPU port of Preparata and
Hong's $O(n\log n)$ 3D divide-and-conquer
algorithm~\cite{Preparata1977ConvexHull}, in the integer-coordinate
formulation of Bullet's
\texttt{btConvexHullComputer}~\cite{Coumans2015Bullet}. All
predicates are integer determinants, so the kernel is free of
floating-point tolerance decisions.  Points are sorted along the
longest bounding-box axis with the warp sort of
Sec.\@~\ref{sec:sort}, then partitioned into $16$ chunks each
assigned to a $2$-lane \emph{hull group} (item 4 in Sec.\@~\ref{sec:warprecipe}). A sub-warp instance of
the model of Sec.\@~\ref{sec:warp}, with one warp hosting $16$
cooperating pairs.  The $16$ sub-hulls merge pairwise in four
balanced-tree rounds.
The same iterative procedure runs in two places: within each 2-lane chunk, and during the final sub-hull merge. At each step, it extends the merged hull by picking the next face — the one that rotates the most around the current bridge edge. The 2-lane group
search from the two bridge endpoints in parallel and resolve by a
single shuffle (item 3 in Sec.\@~\ref{sec:warprecipe}).
Appendix~G of the supplemental illustrates the recursion and the
gift-wrap merge step.

\paragraph{Direction-extreme prefilter.}
For large inputs the D\&C path is dominated by interior points
that cannot appear on the output hull.  The hybrid variant discards
them first: a warp performs $40$ parallel linear scans (item 1 in Sec.\@~\ref{sec:warprecipe}) along the
face-normal directions of a level-$2$ icosphere, retaining $80$
extreme witnesses.  A hull $P_0$ of these witnesses is built;
$P_0$ lies inside the true hull, so any point interior to $P_0$
is non-extreme and dropped, and a final divide-and-conquer call
on the survivors $\cup\;P_0$'s vertices produces the output hull.
On V-HACD this removes ${\sim}70\%$ of points, yielding $>$12$\times$ speedup over the plain D\&C GPU path at $n{=}65,536$ (Sec.\@~\ref{sec:components}); below $n{=}1024$ the prefilter's
overhead exceeds its savings and callers route to the D\&C path
directly.

\paragraph{Hausdorff evaluator.}
\label{sec:hausdorff}
The concavity metric (Eq.\@~\ref{eq:cost})
requires the bidirectional Hausdorff distance $d_H$ between a
part's mesh and its hull.
Like CoACD, we uniformly sample a point set on each mesh, and compute the maximum distance to the triangles on another mesh for $d_H$.
Each mesh--hull pair is assigned single warp.
The whole algorithm can be seen as a traversal on triangles for sampling, and another traversal on sampled points for computing distance, both falling in item 1 of Sec.\@~\ref{sec:warprecipe}.
For distance queries, we employ the LBVH~\cite{Lauterbach2009LBVH} data structure.
Sec.\@~\ref{sec:components} shows performance comparison against CoACD's CPU implementation.

\subsection{Post-processing}
\label{sec:postprocess}

\paragraph{Connected-component splitting.}
\label{sec:cc}
CuACD decomposes any part along its connected components before
adding it to the worklist---once on the input mesh, then on the
sub-parts produced by every accepted cut throughout the search.  We implement this as a warp-parallel
union-find over triangles~\cite{Tarjan1975UnionFind}, with two
adaptations that make it lock-free on the GPU.  First, \textsc{Find} omits path
compression entirely, becoming a purely read-only parent-chain walk;
unions are then the sole writers, and finds never conflict.  Second,
union-by-rank is weakened to best-effort: we attempt the rank
update with a single \texttt{atomicCAS} and skip retries on failure,
since correctness does not depend on its success.  The inner loop is
then a handful of pointer operations with no embedded spin.  The
kernel is lock-free in the strong progress sense (some thread always
makes progress, regardless of scheduling), and on mesh inputs the
forest height remains within the classical $O(\log n)$ envelope.

\paragraph{Hull merge.}
\label{sec:hullmerge}
Once the worklist empties, a greedy merge pass algorithmically
identical to CoACD's combines pairs of output
parts whose joint hull stays below threshold~$\tau$.  On the GPU
each candidate evaluation composes the hybrid convex hull with the
Hausdorff evaluator above, requiring no new geometry kernels, signifying the reusability of our warp-level components.

\section{Results}
\label{sec:results}

All experiments run on an NVIDIA RTX~4090 GPU paired with an
Intel Core~i9-12900K CPU under CUDA~12.1 unless otherwise noted.

\subsection{Implementation Details}
\label{sec:impl}

All hyperparameters are listed in Appendix\@~A; Appendix\@~B ablates them.

We retain
CoACD's~\cite{Wei2022CoACD} threshold default $\tau = 0.05$ for
cross-system comparison, and Fig.\@~\ref{fig:threshold} surveys
CuACD's behavior across a range of $\tau$ values.

The heap allocator reserves $70\%$ of free VRAM as a sticky-slab pool at
context creation.

\subsection{Comparison with Baselines}
\label{sec:comparison}

\paragraph{Datasets and setup.}
We evaluate on three benchmarks: (i) the \emph{V-HACD
benchmark}~\cite{Mamou2016VHACD}, $61$ production meshes used
canonically by CoACD and
NavACD; (ii) a \emph{PartNet-Mobility}
\cite{Xiang2020PartNet} slice of $14{,}085$ per-link merged meshes
across $2{,}347$ articulated objects, a standard throughput stress
test for robotics preprocessing; and (iii) an \emph{Objaverse
subset}~\cite{Deitke2023Objaverse} of $1{,}000$ uniformly sampled
meshes (subsampled because CPU baselines are too slow to run on
the full set; CuACD itself decomposes the full Objaverse without
failure), probing in-the-wild geometry beyond curated benchmarks.  We compare against NavACD~\cite{Lands2024NavACD},
VisACD~\cite{Fokin2026VisACD}, and CoACD~\cite{Wei2022CoACD}, sweeping
each baseline's concavity threshold until its mean output concavity
matches $\tau = 0.05$ on V-HACD; at this matched-quality operating point we
report mean per-mesh decomposition time and output part count.
All methods consume the same PaMO~\cite{Oh2025PaMO}-preprocessed
inputs at ratio $0.1$ with $10{,}000$ minimum faces to remove
preprocessing variance. We exclude the PaMO stage from all timings.  Across every benchmark and
ablation, CuACD completes every input with zero failures, including
meshes with degenerate triangles.  Behavior up to million-triangle
inputs is reported in Sec.\@~\ref{sec:scaling}.

We exclude the contemporary learning-based
RL-ACD~\cite{Luo2025RLACD} and~\citet{Knodt2026CPD}, as neither
releases code, weights, or training data.

\begin{figure*}[!t]
  \centering
  \includegraphics[width=0.96\linewidth]{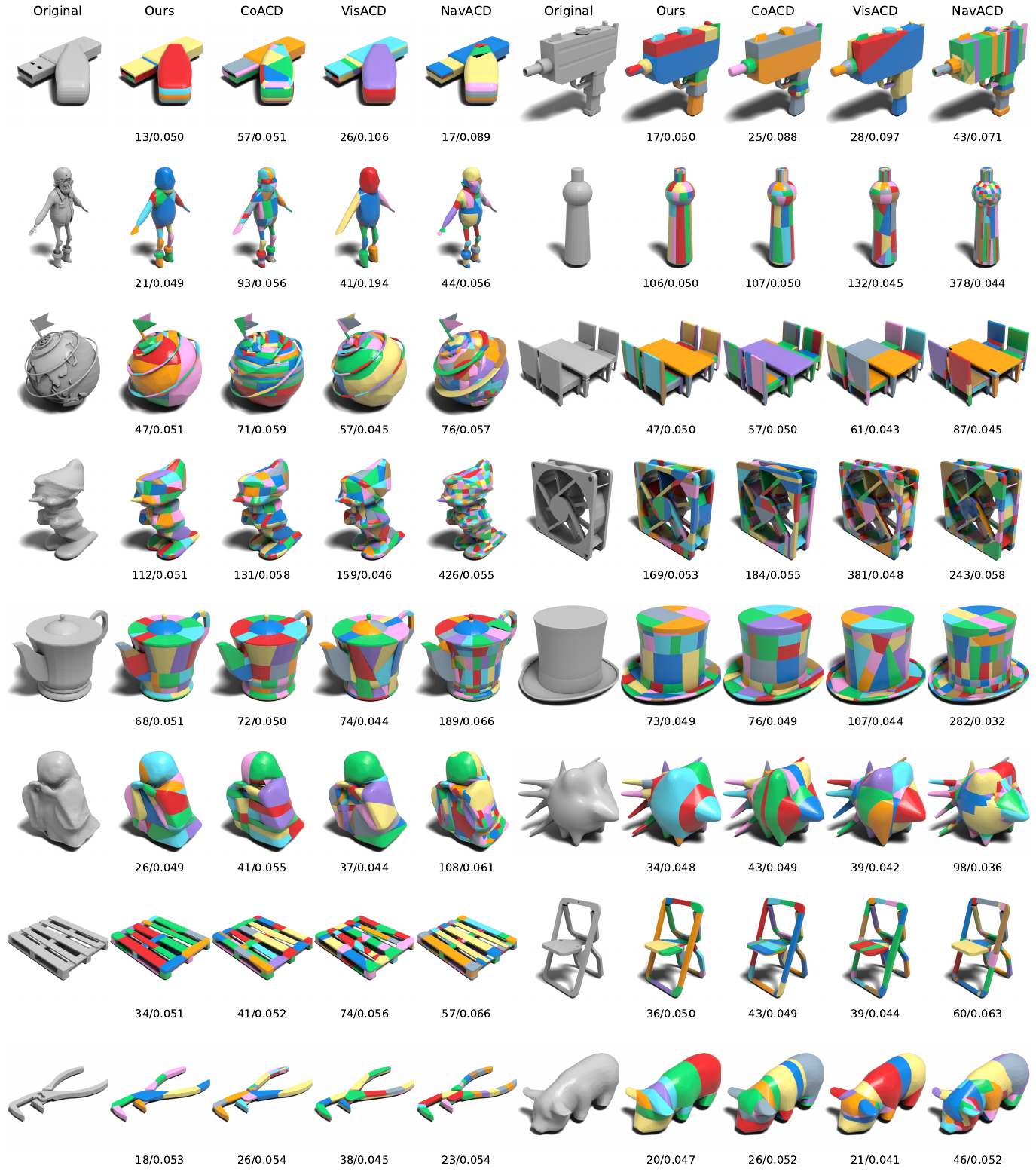}
  \caption{Visual comparison of decompositions produced by CoACD,
    NavACD, VisACD, and CuACD at matched mean concavity on
    representative meshes from the V-HACD and PartNet-Mobility
    benchmarks.  The two numbers under each decomposition report
    \emph{\#parts\,/\,max per-part concavity}.}
  \label{fig:qualitative}
\end{figure*}

\paragraph{Quantitative analysis.}
From Table~\ref{tab:comparison}, CuACD is $\mathbf{78\times}$, $\mathbf{80\times}$, and
$\mathbf{104\times}$ faster than CoACD on V-HACD, PartNet-Mobility, and the
Objaverse subset respectively, at matched or lower mean concavity
and comparable or fewer parts; the gap over VisACD---the fastest
prior GPU-assisted baseline---is $40$--$64\times$.  Even on a laptop RTX 3080 Mobile, our method averages $0.64$\,s per mesh on the V-HACD benchmark — an order of magnitude faster than every CPU baseline.
\paragraph{Qualitative comparison.}
Figure~\ref{fig:qualitative} shows CuACD aligning its cuts with
semantically meaningful features (handles, wheels, articulated joints)
on par with or cleaner than CPU baselines despite running orders of
magnitude faster, thanks to searching for more candidates than the CPU
baselines can afford.

\subsection{GPU Gains Versus Algorithmic Gains}
\label{sec:attribution}

Table~\ref{tab:comparison} mixes two effects---the GPU implementation
and the enriched candidate pool---and we separate them here.

\paragraph{Candidate generation.}
CuACD outperforms CoACD in quality at hyperparameters that are
nominally weaker ($D=2$ rather than $4$, $W_{\mathrm{sub}}=5$ rather
than $60$). This is because the \emph{root} candidate pool is enriched with $4N_{\mathrm{ce}}=64$ concave-edge planes to a total $94$ root candidates, which is broader than CoACD's $60$.  At a fixed threshold the
concave-edge family brings the part count down, while we observe deeper branching
has minimal impact. The full ablation results are detailed in supplemental B.

\paragraph{A matched CPU implementation.}
To further understand our system, we ported CuACD's concave edge sampling and hyperparameters to CoACD. The algorithm now mostly matches except the exploration algorithm in the search is still guided by MCTS and not a look-ahead tree search.
We ran this modifed version of CoACD on the same V-HACD benchmark we mentioned before. This CPU port averages $33.7$ parts and $0.0481$
maximum concavity, largely matching CuACD's $33.6$ and $0.0488$. It runs at $20.85$\,s
per mesh, compared to original CoACD's $18.03$ and CuACD's $0.23$\,s.  The quality gain over stock CoACD
is therefore the enriched pool and is reproducible on the CPU. On the other hand, this also proves that CuACD's
speedup mostly comes from our system design that runs the algorithm efficiently on the GPU.

\paragraph{GPU profile.}
The main kernels sustain $25$--$50\%$ warp occupancy, bounded by the
registers held by intermediate algorithmic state.%
\subsection{Reusable Component Performance}
\label{sec:components}

\begin{figure}[t]
  \centering
  \includegraphics[width=\linewidth]{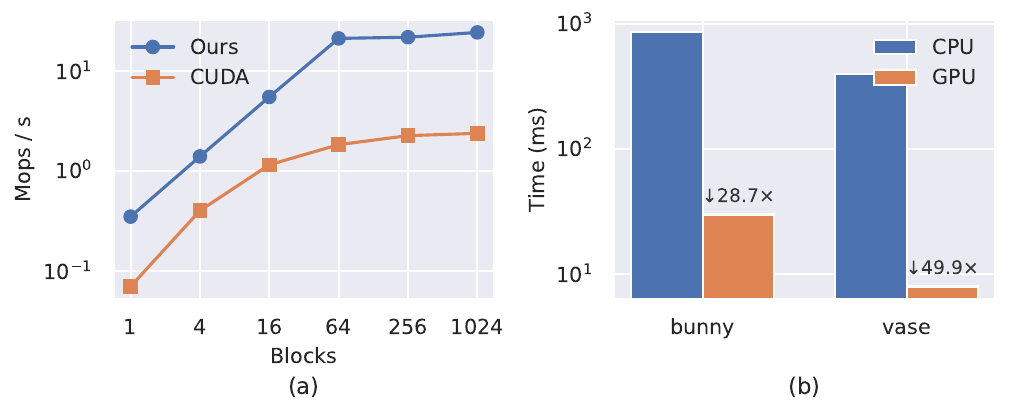}
  \vspace{-2em}
  \caption{\textbf{Left:} allocation throughput w.r.t number of active warps on the stress
    test in log-log scale. Higher is better.
    \textbf{Right:} bidirectional Hausdorff distance timing for a
    batch of $100$ mesh--hull pairs compared against the CoACD CPU reference~\cite{Wei2022CoACD};
    lower is better.}
  \label{fig:malloc-hausdorff}
\vspace{-1em}
\end{figure}

\begin{figure}[!ht]
  \setlength{\abovecaptionskip}{2pt}
  \setlength{\belowcaptionskip}{0pt}
  \centering
  \includegraphics[width=\linewidth]{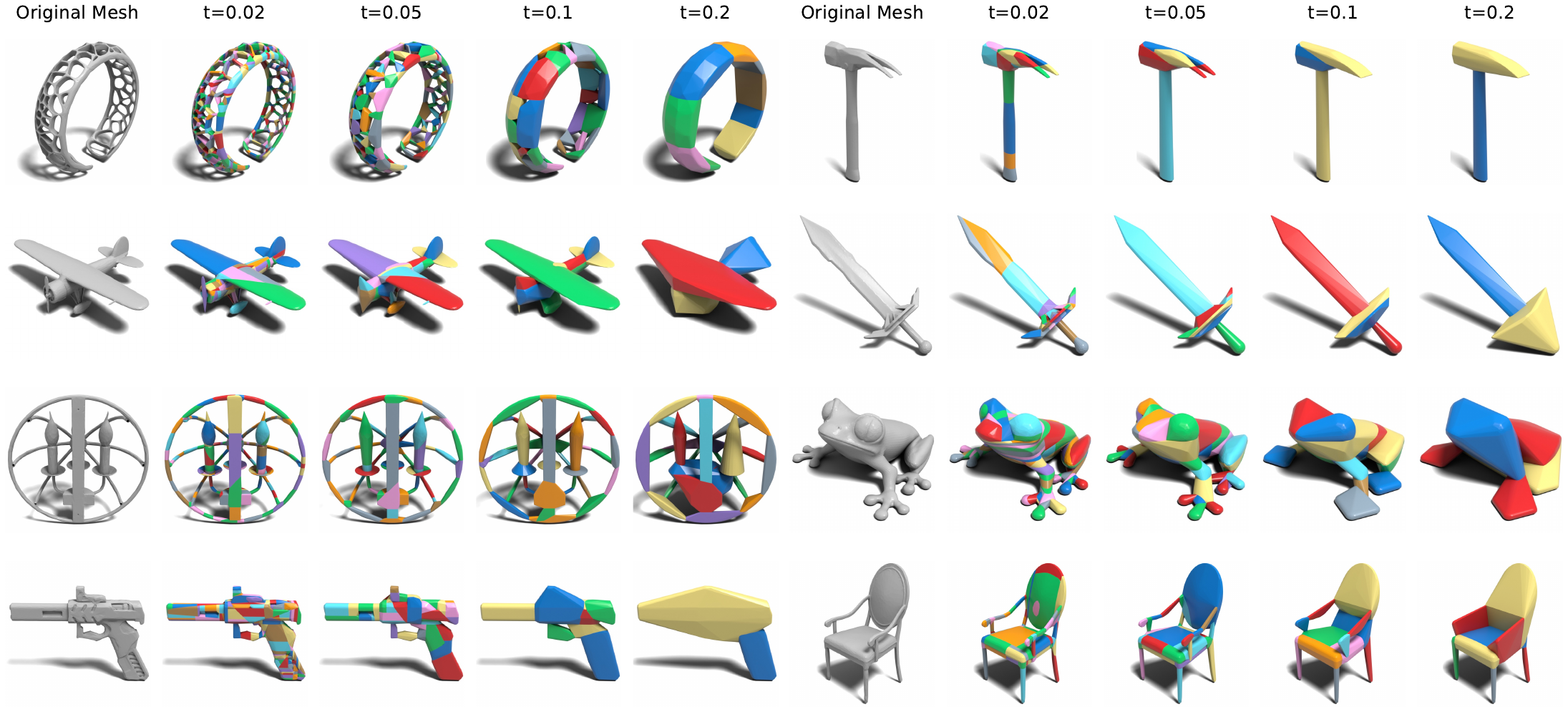}
  \caption{CuACD decompositions across a range of concavity
    thresholds~$\tau$.  Tighter thresholds preserve finer geometric
    features at the cost of more parts; looser thresholds yield
    coarser decompositions better suited to fast collision queries.
    Across the full sweep, cuts remain aligned with semantically
    meaningful boundaries, and runtimes stay within the order
    reported in Table~\ref{tab:comparison}.}
  \vspace{-1em}

  \label{fig:threshold}
\end{figure}

\paragraph{Device-side heap allocator.}
Figure~\ref{fig:malloc-hausdorff} (left) compares our allocator
against CUDA's \texttt{malloc}/\texttt{free} on a glibc
\texttt{mstress}-style stress test ($1$--$1024$ blocks, log-uniform
sizes in $[4\,\mathrm{KiB}, 256\,\mathrm{KiB}]$): ours scales linearly to the configurable arena count, which is set to 64 in the experiment, and leads at
every grid size, $5\times$ contention-free and $10$--$11\times$
when arenas are oversubscribed.

\begin{table}[t]
\caption{Convex hull construction time on batches of $256$
  independent point sets, in milliseconds. Within each row, bold
  marks the faster GPU method and underline the second.}
\label{tab:hull}
\centering
\small
\begin{tabular}{llccc}
  \toprule
  \textbf{Distribution} & \textbf{$n$}
    & \textbf{D\&C CPU} & \textbf{D\&C GPU} & \textbf{Hybrid} \\
  \midrule
  \multirow{4}{*}{Uniform cube}
    & $512$      & $47.73$   & $\underline{1.66}$    & $\mathbf{1.65}$  \\
    & $2{,}048$  & $198.23$  & $\underline{6.47}$    & $\mathbf{1.33}$  \\
    & $16{,}384$ & $1729.65$ & $\underline{59.77}$   & $\mathbf{5.00}$  \\
    & $65{,}536$ & $6839.78$ & $\underline{260.76}$  & $\mathbf{20.53}$ \\
  \midrule
  \multirow{4}{*}{Gaussian}
    & $512$      & $44.33$   & $\underline{1.60}$    & $\mathbf{1.58}$  \\
    & $2{,}048$  & $188.42$  & $\underline{6.35}$    & $\mathbf{0.89}$  \\
    & $16{,}384$ & $1619.98$ & $\underline{60.04}$   & $\mathbf{3.33}$  \\
    & $65{,}536$ & $5561.01$ & $\underline{243.08}$  & $\mathbf{17.61}$ \\
  \bottomrule
\end{tabular}
\vspace{-1.5em}
\end{table}
\paragraph{Convex hull.}
Table~\ref{tab:hull} benchmarks batches of $256$ point sets
($n{=}512$ to $65{,}536$) against Bullet's CPU
hull~\cite{Coumans2015Bullet}: the hybrid variant runs
$>$300$\times$ faster than Bullet and $>$12$\times$ faster than the
plain D\&C GPU path at $n{=}65{,}536$.

\paragraph{Hausdorff distance.}
Figure~\ref{fig:malloc-hausdorff} (b) reports our bidirectional
Hausdorff timing on \emph{bunny} and \emph{vase} batches against
the CoACD reference~\cite{Wei2022CoACD} (also used by VisACD).

\paragraph{Sorting.}
Our warp-cooperative quick-sort scales beyond $10^{3}$ elements
where CUB's kernels do not. See Appendix~C for more details.

\subsection{Scalability and Threshold Sensitivity}
\label{sec:scaling}

\begin{figure}[!ht]
  \setlength{\abovecaptionskip}{2pt}
  \setlength{\belowcaptionskip}{0pt}
  \centering
  \includegraphics[width=\linewidth]{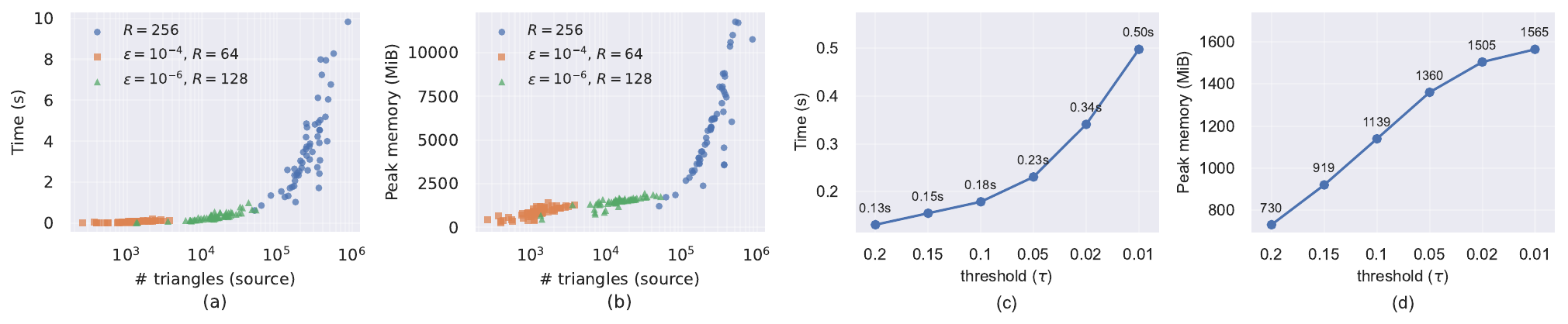}
  \caption{CuACD scalability on the V-HACD benchmark.
    \textbf{(a)}~End-to-end decomposition time and \textbf{(b)}~peak
    GPU memory versus the source mesh's triangle count, across PaMO
    preprocessing settings parameterized by the curvature-related
    simplification terminal threshold $\varepsilon$ and the remesh
    resolution $R$; the no-simplification ``$R{=}256$'' setting
    functions as a stress test well above the resolutions used by
    VisACD ($R{=}20$) and CoACD ($R{=}50$).
    \textbf{(c)}~Mean per-mesh time and \textbf{(d)}~mean peak GPU
    memory as a function of the concavity threshold $\tau$, averaged
    at the default PaMO setting; each marker is annotated with its
    value.}
  \vspace{-1.5em}

  \label{fig:scaling}
\end{figure}

Beyond the PaMO~\cite{Oh2025PaMO} preprocessing setting used in
Sec.\@~\ref{sec:comparison},
Figure~\ref{fig:scaling}(a,b) stress-tests CuACD on substantially
heavier inputs.  We sweep three PaMO settings of $(\varepsilon, R)$ --- the simplification tolerance and remesh resolution --- from coarse to fine: $(10^{-4}, 64)$, $(10^{-6}, 128)$, and $(\text{none}, 256)$, with the finest reaching ${\sim}10^{6}$ triangles.  Million-triangle inputs are not a realistic
ACD use case, but CuACD's runtime grows gracefully into the
multi-second regime even at $10^{6}$ triangles, and peak GPU memory
stays comfortably within a single RTX~4090's $24$\,GiB envelope.

Figure~\ref{fig:scaling}(c,d) sweeps the concavity threshold $\tau$
from $0.2$ down to $0.01$. At the default $\tau{=}0.05$ CuACD averages
$0.23$\,s per mesh; tightening to $\tau{=}0.01$ raises the average
part count from ${\sim}34$ to ${\sim}194$ and the runtime to
${\sim}0.5$\,s, while peak memory grows only from $1.4$ to
$1.6$\,GiB---an order-of-magnitude finer decomposition for roughly
$2\times$ wall-clock cost.


\subsection{Application: Scene-Scale Decomposition}
\label{sec:scene}
\begin{figure}[!ht]
  \setlength{\abovecaptionskip}{8pt}
  \setlength{\belowcaptionskip}{0pt}
  \centering
  \includegraphics[width=\linewidth]{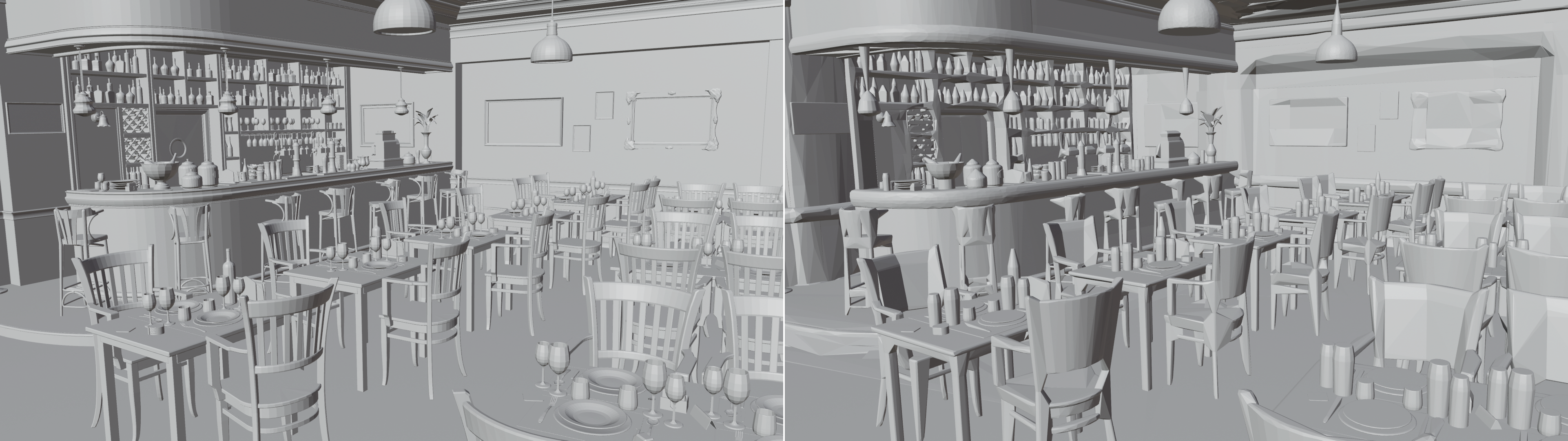}
  \caption{\textbf{Left:} input mesh. \textbf{Right:} CuACD output.
    CuACD applied to the Amazon Lumberyard
    Bistro~\cite{Lumberyard2017Bistro} interior ($585{,}299$
    vertices, $1{,}043{,}077$ triangles): at $\tau{=}0.005$, CuACD
    decomposes the full scene in $18.73$ seconds with $2355.6$\,MiB
    peak GPU memory on a single RTX~4090.}
  \vspace{-1.5em}

  \label{fig:scene}
\end{figure}

The most operationally meaningful test of CuACD is to run it on a
whole scene rather than one mesh at a time.
Figure~\ref{fig:scene} shows CuACD applied to the Amazon
Lumberyard Bistro~\cite{Lumberyard2017Bistro} interior
($585{,}299$ vertices, $1{,}043{,}077$ triangles): at
$\tau{=}0.005$, CuACD decomposes the full scene in $18.73$\,s with
$2355.6$\,MiB peak GPU memory on a single RTX~4090.

\section{Conclusion and Discussion}
\label{sec:discussion}

CuACD adapts the warp-as-unit view of \citet{Hong2011WarpCentric}
from graph traversal to computational geometry, delivering the first
fully GPU-resident ACD orders of magnitude faster than CoACD at
matched or better quality across three benchmarks.

\paragraph{Limitations and future work.}
CuACD retains the hand-designed search heuristics of its CoACD lineage
and does not use learned proposals; some cuts are suboptimal relative
to a trained policy, and we leave learned cut quality to future work.
The device-side heap does not compact across arenas after
fragmentation, nor return freed memory to the system allocator; the
main kernels sustain $25$--$50\%$ SM occupancy, bounded by register
pressure.  Combining the warp-cooperative allocation path with modern
memory management and register-aware scheduling is future work.%

\bibliographystyle{ACM-Reference-Format}
\bibliography{sample-bibliography}

\newpage
\appendix

\section{Hyperparameter Reference}
\label{app:params}

Table~\ref{tab:params} lists every hyperparameter exposed by CuACD,
grouped by role.  Two parameters ($\tau$ and $k$) define the concavity
metric; two ($\varepsilon_{\mathrm{axis}}$ and
$\varepsilon_{\mathrm{conc}}$) set geometric tolerances on plane
placement; three ($W_{\mathrm{top}}$, $W_{\mathrm{sub}}$, $D$) govern
the shape of the look-ahead tree; $N_{\mathrm{ce}}$ bounds the
concave-edge proposal budget per part; and $I_{\mathrm{ce}}$ controls
for how many outer iterations the concave-edge proposal family
is active.  We retain the CoACD threshold default $\tau = 0.05$
so that concavity numbers are directly comparable across systems.

\begin{table}[t]
\caption{CuACD hyperparameters and the default values used throughout
  the paper unless otherwise noted.}
\label{tab:params}
\begin{center}
\begin{tabular}{llc}
  \toprule
  \textbf{Role} & \textbf{Symbol} & \textbf{Value} \\
  \midrule
  Concavity threshold                 & $\tau$                         & $0.05$ \\
  Concavity calibration coefficient   & $k$                            & $0.3$ \\
  Axis-plane boundary margin          & $\varepsilon_{\mathrm{axis}}$  & $\min(\tau/4,\,0.015)$ \\
  Concave-edge plane offset           & $\varepsilon_{\mathrm{conc}}$  & $0.005$ \\
  Axis-aligned proposals at the root  & $W_{\mathrm{top}}$             & $30$ \\
  Deeper-level branching factor       & $W_{\mathrm{sub}}$             & $5$ \\
  Look-ahead depth                    & $D$                            & $2$ \\
  Sampled concave edges per part      & $N_{\mathrm{ce}}$              & $16$ \\
  Concave-edge iteration budget       & $I_{\mathrm{ce}}$              & $10$ \\
  \bottomrule
\end{tabular}
\end{center}
\end{table}

\section{Ablation of Search Hyperparameters}
\label{app:ablation}

\begin{figure}[t]
  \centering
  \includegraphics[width=\linewidth]{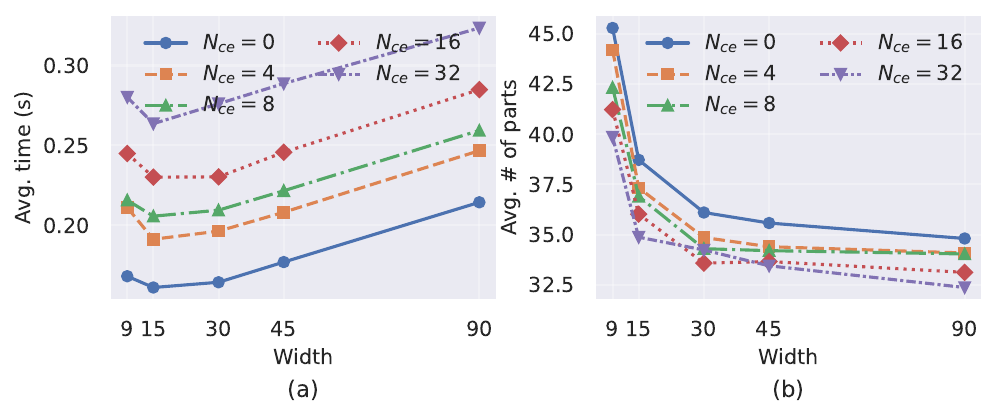}
    \vspace{-2em}
  \caption{Ablation of the two root-branching hyperparameters on
    the V-HACD benchmark.  \textbf{(a)}~Mean per-mesh decomposition
    time.  \textbf{(b)}~Mean output part count.  Each curve fixes
    $N_{\mathrm{ce}}$ and sweeps $W_{\mathrm{top}}$; the chosen
    default is $(W_{\mathrm{top}}=30,\,N_{\mathrm{ce}}=16)$.}
  \label{fig:ablation}
\end{figure}

\paragraph{Root branching factors.}
The two parameters that govern the size of the root candidate pool
are $W_{\mathrm{top}}$, the number of axis-aligned proposals, and
$N_{\mathrm{ce}}$, the number of sampled concave edges (each emitting
four planes), for a total root branching factor of
$W_{\mathrm{top}} + 4\,N_{\mathrm{ce}}$.
Figure~\ref{fig:ablation} sweeps
$W_{\mathrm{top}} \in \{9, 15, 30, 45, 90\}$ at five fixed settings
of $N_{\mathrm{ce}}$ on the V-HACD benchmark, reporting both mean
per-mesh runtime and mean output part count.  Two regimes are
visible.  Below the knee at
$(W_{\mathrm{top}}{=}30,\,N_{\mathrm{ce}}{=}16)$, undersized pools
miss high-quality cuts on geometrically intricate parts and the
search compensates with extra outer iterations, inflating the part
count even though each individual evaluation is cheap; the curves
in panel (b) climb steeply as the pool shrinks.  Above the knee, the
part count plateaus---additional candidates are essentially
duplicates of cuts already proposed---while runtime continues to grow
roughly linearly with the total branching factor (panel a), since
each candidate triggers a full bounded look-ahead evaluation.  The
two knees coincide closely across the family of curves, indicating
that $W_{\mathrm{top}}$ and $N_{\mathrm{ce}}$ contribute largely
independent coverage of the cut space rather than redundantly
proposing the same planes.  The default
$(W_{\mathrm{top}}{=}30,\,N_{\mathrm{ce}}{=}16)$ sits within noise of
the quality plateau at a small fraction of the cost incurred by
larger pools, and we use it throughout the paper.

\paragraph{Remaining knobs.}
Table~\ref{tab:ablation-misc} covers the remaining search knobs.
Running CC-Split after every cut rather than only at the end reduces
the part count by roughly seven and slightly shortens runtime.
Reducing $I_{\mathrm{ce}}$ from $10$ to $3$ shaves ${\sim}17\%$ off
the runtime at a cost of ${\sim}2$ extra parts; raising it to $20$
yields no measurable improvement.  $W_{\mathrm{sub}}=3$ matches the
default within noise, while $W_{\mathrm{sub}}=15$ roughly triples the
look-ahead leaf count for under one part of gain, confirming
$W_{\mathrm{sub}}=5$ as a clear operating point.  Deepening the
look-ahead from $D=2$ to $D=3$ leaves the mean part count essentially
unchanged ($32.89$ vs.\ $33.57$) while nearly doubling runtime
($0.41$~s vs.\ $0.23$~s), confirming depth $2$ as the sweet spot.

\begin{table}[t]
\caption{Ablation of the remaining search knobs on V-HACD, varying
  one setting at a time from the defaults of
  Table~\ref{tab:params}.  Lower is better on both axes.}
\label{tab:ablation-misc}
\begin{center}
\small
\begin{tabular}{lcc}
  \toprule
  \textbf{Setting} & \textbf{Avg.\ time (s)} & \textbf{Avg.\ \#parts} \\
  \midrule
  Default                        & $0.23$ & $33.57$ \\
  Look-ahead depth $D = 3$ & $0.41$ & $32.89$ \\
  No per-iter CC-Split           & $0.25$ & $40.87$ \\
  $I_{\mathrm{ce}} = 3$          & $0.19$ & $35.20$ \\
  $I_{\mathrm{ce}} = 20$         & $0.23$ & $33.54$ \\
  $W_{\mathrm{sub}} = 3$         & $0.23$ & $33.44$ \\
  $W_{\mathrm{sub}} = 15$        & $0.44$ & $33.07$ \\
  \bottomrule
\end{tabular}
\end{center}
\end{table}

\paragraph{Per-mesh distribution.}
The averages above and in the main paper's Table~1 aggregate the
$61$-mesh V-HACD benchmark, whose per-mesh outcomes vary with input
geometry.  Table~\ref{tab:vhacd-quantiles} reports the full
distribution for CuACD.  Part counts span more than an order of
magnitude ($3$--$165$), reflecting how some inputs decompose into a
handful of near-convex pieces while others fragment heavily.  Runtimes
track the part count but stay sub-half-second on every mesh, and
maximum concavity never exceeds the threshold $\tau=0.05$ by more than
sampling error, since CuACD accepts a part only once its measured
concavity is at or below $\tau$.  Because the spread is driven entirely
by input geometry rather than by CuACD's settings, isolated per-mesh
metrics are uninformative on their own; the matched-mesh means in
Table~1 remain the right basis for cross-system comparison.

\begin{table}[t]
\caption{CuACD per-mesh distribution on the $61$-mesh V-HACD
  benchmark: minimum, first quartile (Q1), median, third quartile (Q3),
  and maximum.}
\label{tab:vhacd-quantiles}
\begin{center}
\small
\begin{tabular}{lccccc}
  \toprule
  \textbf{Metric} & \textbf{Min} & \textbf{Q1} & \textbf{Median} & \textbf{Q3} & \textbf{Max} \\
  \midrule
  Output parts       & $3$    & $16$    & $24$    & $43$    & $165$  \\
  Runtime (s)        & $0.06$ & $0.18$  & $0.22$  & $0.29$  & $0.44$ \\
  Max concavity      & $0.023$ & $0.0477$ & $0.0497$ & $0.0513$ & $0.059$ \\
  \bottomrule
\end{tabular}
\end{center}
\end{table}

\section{Warp-Cooperative Sort: Component Benchmark}
\label{app:sort}

Table~\ref{tab:sort} compares the warp-cooperative quick-sort
described in Sec.\@~3.1 of the main paper against
\texttt{std::sort} on the CPU and CUB's block-scope merge and radix
sorts, on batches of $1{,}000$ \texttt{int4} sequences from $10^{2}$
to $10^{5}$ elements.  Our kernel is the only GPU method that
remains feasible at the largest sizes---CUB's block functions exceed
shared-memory capacity beyond $10^{3}$---while requiring only $24$
bytes of shared memory and accepting dynamically sized inputs.  At
the sizes where CUB merge is feasible, our kernel stays within a
small constant factor of it.

\begin{table}[t]
\caption{Sort throughput on batches of $1{,}000$ independent sequences
  of \texttt{int4} keys.  Bold marks the best time and shared-memory
  requirement; underline marks the second best among GPU methods.}
\label{tab:sort}
\begin{center}
\small
\begin{tabular}{llcc}
  \toprule
  \textbf{Seq.\ len.} & \textbf{Method} & \textbf{Time (ms)} & \textbf{Smem (B)} \\
  \midrule
  \multirow{4}{*}{$10^{2}$}
    & \texttt{std::sort}  & $1.5748$               & ---           \\
    & CUB merge           & $\mathbf{0.0070}$      & $2064$        \\
    & CUB radix           & $0.0417$               & $4640$        \\
    & Ours                & $\underline{0.0135}$   & $\mathbf{24}$ \\
  \midrule
  \multirow{4}{*}{$10^{3}$}
    & \texttt{std::sort}  & $25.0594$              & ---           \\
    & CUB merge           & $\mathbf{0.0667}$      & $16400$       \\
    & CUB radix           & $0.1665$               & $16896$       \\
    & Ours                & $\underline{0.1579}$   & $\mathbf{24}$ \\
  \midrule
  \multirow{3}{*}{$10^{4}$}
    & \texttt{std::sort}     & $326.9430$         & ---           \\
    & CUB merge / radix      & \emph{fail}        & ---           \\
    & Ours                   & $\mathbf{4.5052}$  & $\mathbf{24}$ \\
  \midrule
  \multirow{3}{*}{$10^{5}$}
    & \texttt{std::sort}     & $4062.6499$        & ---           \\
    & CUB merge / radix      & \emph{fail}        & ---           \\
    & Ours                   & $\mathbf{71.3206}$ & $\mathbf{24}$ \\
  \bottomrule
\end{tabular}
\end{center}
\end{table}

\section{Warp-Cooperative Sort: Algorithm and Mechanics}
\label{app:sort-mechanics}

This appendix gives the explicit algorithm and instruction sequence
of the warp-cooperative quick-sort summarized in Sec.\@~3.1 of the
main paper.

\begin{algorithm}[t]
  \caption{Warp-cooperative quick-sort (per warp).}
  \label{alg:wcsort-supp}
  \begin{algorithmic}[1]
    \Require segment $A[l..r]$; lane id $\ell\!\in\![0,31]$
    \State push $(l, r)$ onto warp-private stack
    \While{stack non-empty}
      \State $(l, r) \gets \text{pop}$
      \If{$r - l \leq 32$}
        \State \Call{BitonicSort}{$A[l..r]$} \Comment{base case}
      \Else
        \State $p \gets \Call{Median32}{A[l..r]}$
        \State $(m_1, m_2) \gets \Call{WarpPartition3}{A[l..r], p}$
        \State push $(m_2{+}1, r)$ and $(l, m_1{-}1)$
      \EndIf
    \EndWhile
  \end{algorithmic}
\end{algorithm}

\paragraph{Warp-parallel three-way partition.}
Given a pivot $p$, we sweep the segment in $32$-element chunks with
a 1-to-1 chunk-to-lane assignment.  Lane~$\ell$ loads one element
$e$ and contributes to two warp-wide masks via
\texttt{\_\_ballot\_sync}: $M_{<}$ collects $e<p$, $M_{>}$ collects
$e>p$.  Each lane's output slot in the corresponding region is the
prefix popcount
$\texttt{\_\_popc}\!\bigl(M_{*}\,\&\,((1\!\ll\!\ell)-1)\bigr)$
over lower-numbered lanes, so the $32$ lanes commit to unique
destinations concurrently with no inter-lane synchronization beyond
the implicit barrier of the ballot.  Cooperation is expressed
entirely in two warp-wide votes and a prefix popcount.  Less-than
elements grow the left region forward and greater-than elements
grow the right region backward; the interior is then filled with
$p$, yielding $(<\!p\,|\,=\!p\,|\,>\!p)$ in $O(n)$ loads with zero
auxiliary shared memory.

\begin{figure}[t]
  \centering
  \includegraphics[width=\linewidth]{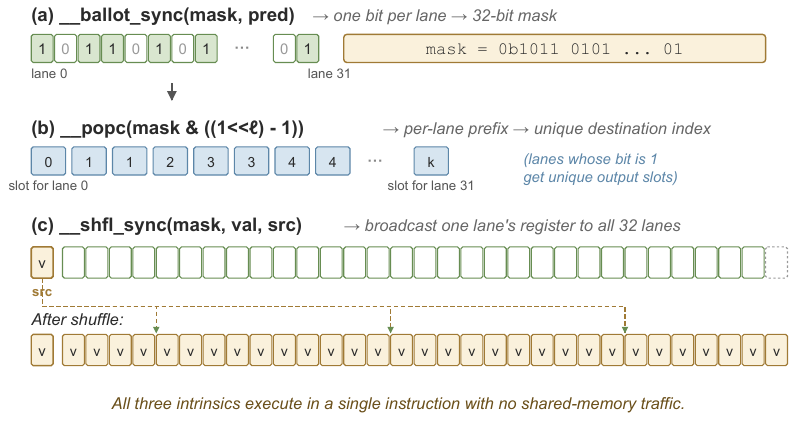}
    \vspace{-2em}
  \caption{Three warp-level intrinsics carrying the partition of
    Algorithm~\ref{alg:wcsort-supp}.  \textbf{Top:}
    \texttt{\_\_ballot\_sync} collects one predicate bit from each
    of the $32$ lanes into a single $32$-bit mask.
    \textbf{Middle:} \texttt{\_\_popc} of the mask AND-ed with
    $((1\!\ll\!\ell)-1)$ gives lane $\ell$ a unique prefix index
    among the lanes whose predicate fired.  \textbf{Bottom:}
    \texttt{\_\_shfl\_sync} delivers a register value from a source
    lane to every lane in the warp in one instruction.}
  \label{fig:warp-primitives}
\end{figure}

\paragraph{Primitive operations.}
Three CUDA warp-level intrinsics carry the partition
(Figure~\ref{fig:warp-primitives}):
\texttt{\_\_ballot\_sync(mask,\,pred)} returns a 32-bit integer in
which bit $\ell$ is set iff lane $\ell$'s predicate is true;
\texttt{\_\_popc(x)} returns the population count of an integer;
and \texttt{\_\_shfl\_sync} broadcasts a register value from a
source lane to every lane in the warp.
All three execute in a single instruction with no shared-memory
traffic.

\section{Heap Allocator Internals}
\label{app:heap-internals}

This appendix details the warp-level entry, slab sizing, and
boundary-tag coalescing summarized by the caption of Figure~3 in
the main paper and Sec.\@~3.3.

\paragraph{Spin-lock and boundary tags.}
A \emph{spin-lock} is a lock acquired by repeatedly polling a
memory location until it is observed unlocked.  Only lane~$0$ of
each warp holds the per-arena lock, so the polling traffic is
bounded by the number of co-resident warps sharing the arena rather
than the raw thread count.  Each block stores a small header at
both ends---a \emph{boundary tag} in the sense of
Knuth~\cite{Knuth1997TAOCP1}---so on free we locate the immediate
neighbor blocks in $O(1)$ pointer chases and merge them if they too
are free.  No offline compaction pass is ever required.

\paragraph{$O(1)$ bin lookup and slab sizing.}
Each arena maintains $32$ power-of-two bins, each split at its
midpoint into two sub-bins, so a single $64$-bit bitmap records
which sub-bins are non-empty.  The smallest feasible sub-bin is
located in one \texttt{\_\_ffsll} instruction (find-first-set on a
64-bit integer) and its free-list head is popped directly, making
an in-bin allocation a constant-time operation under the spin-lock.
When no bin is feasible the allocator bumps the underlying linear
pool for a fresh slab sized to at least twice the aligned request
plus sentinel overhead.  This invariant guarantees that the
post-split remainder lands in a strictly higher sub-bin than the
current request, so the next request of the same shape hits the
free list rather than the pool, and the pool size stabilizes after
the first decomposition call.

\section{Mesh--Plane Clipper Internals}
\label{app:clipper-internals}

This appendix details the ear-clipping sub-procedure inside the
mesh--plane clipper.  The surrounding pipeline---sign classification,
triangle splitting, boundary discovery, and loop reconstruction---is
already covered in Sec.\@~4.2 and Algorithm~1 of the main paper and
is not restated here.

\paragraph{Ears.}
Each open boundary loop is closed by ear-clipping.  Outer and inner
loops are first classified by 2D ray casting and inner loops are
bridged into their enclosing outer loop by a single chord, after
which the cap becomes a simple polygon.  An \emph{ear} is a vertex
$v_{\rm cur}$ whose polygon neighbors $v_p$ and $v_n$ satisfy two
conditions: (i)~$v_{\rm cur}$ is convex---the cross product of the
incoming and outgoing edges has the same sign as the polygon's
orientation---and (ii)~no other polygon vertex lies inside the
triangle $(v_p, v_{\rm cur}, v_n)$.  Clipping an ear emits that
triangle, removes $v_{\rm cur}$ from the polygon, and reduces the
vertex count by one.  Iterating until three vertices remain yields a
complete triangulation.

\begin{figure}[t]
  \centering
  \includegraphics[width=\linewidth]{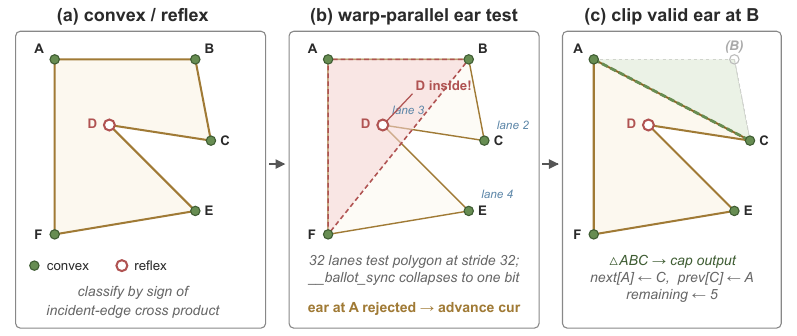}
    \vspace{-2.5em}
  \caption{Warp-cooperative ear-clipping on a cap loop.
    \textbf{(a)}~Convex and reflex vertices are classified by the
    sign of the cross product of incident edges; D is the only
    reflex vertex, so A, B, C, E, F are ear candidates.
    \textbf{(b)}~For the candidate ear at A, the 32 warp lanes scan
    polygon indices at stride 32 and test in parallel whether any
    other vertex lies inside triangle $(F, A, B)$; D is inside, so
    \texttt{\_\_ballot\_sync} rejects the ear and the cursor
    advances.  \textbf{(c)}~At B the test passes; triangle $ABC$ is
    written to the cap output, B is spliced out of the
    \texttt{prev}/\texttt{next} list, and the new diagonal $AC$
    becomes a polygon edge.}
  \label{fig:ear-clip}
\end{figure}

\paragraph{Warp-cooperative ear test.}
The polygon is held in shared memory as a doubly-linked list with
arrays \texttt{prev} and \texttt{next} (case~(c) of the recipe in
App.~I).  Lane~0 of the cap warp walks the list with a cursor
\texttt{cur} (Figure~\ref{fig:ear-clip}).  At each step lane~0 reads
$v_p \!=\! \texttt{prev[cur]}$ and $v_n \!=\! \texttt{next[cur]}$,
computes the cross product, and broadcasts the result via
\texttt{\_\_shfl\_sync}.  If $v_{\rm cur}$ is reflex, the cursor
advances; if $|\text{cross}| \!\le\! \varepsilon$, a degenerate
zero-area ear is emitted to keep the cap watertight at T-junctions
and along edges split into collinear segments.  Otherwise the warp
runs a parallel point-in-triangle test: each of the 32 lanes scans
polygon indices at stride 32 and tests its assigned vertices against
the candidate ear, and \texttt{\_\_ballot\_sync} collapses the
per-lane hits into a single warp-wide bit (case~(a) of the recipe).
If no hit, $v_{\rm cur}$ is an ear; lane~0 emits the triangle,
splices $v_{\rm cur}$ out of the list by writing
\texttt{next[}$v_p$\texttt{]}$\!\gets\!v_n$ and
\texttt{prev[}$v_n$\texttt{]}$\!\gets\!v_p$, and decrements the
remaining count.  Cost is $O(N)$ work per ear in $O(N/32)$ rounds,
$O(N^2/32)$ across the polygon.

\section{Convex Hull Internals}
\label{app:hull-internals}

This appendix details the divide-and-conquer convex hull
summarized in Sec.\@~4.3.

\begin{figure}[t]
  \centering
  \includegraphics[width=\linewidth]{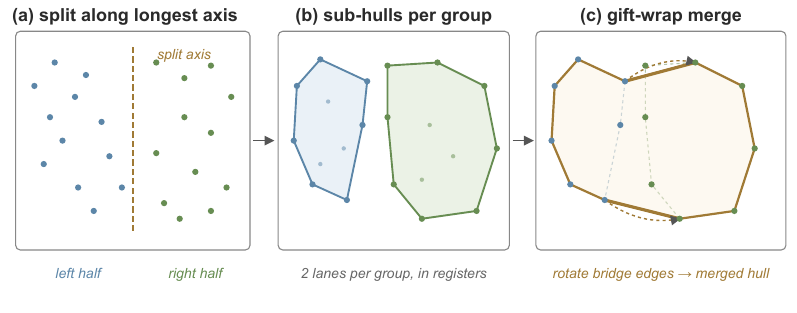}
    \vspace{-2.5em}
  \caption{One step of the divide-and-conquer convex hull.
    \textbf{Left:} the input point set is split along the longest
    bounding-box axis into two halves.  \textbf{Middle:} each half
    is hulled independently inside a hull group.  \textbf{Right:}
    the two child hulls are merged by gift-wrapping the bridge edge
    around their boundaries.}
  \label{fig:hull-dnc}
\end{figure}

\paragraph{Hull-pair merge.}
The merge step in the divide-and-conquer hull of Sec.\@~4.3 uses
gift-wrapping~\cite{Preparata1977ConvexHull}: it picks an initial
cross-edge spanning the two extreme points along the merge axis,
then rotates that edge around its endpoints, at each step
advancing whichever endpoint exposes the larger turning angle,
until the rotation closes a ring.  The two lanes of each hull
group search from the two bridge endpoints in parallel and resolve
each step with a single warp shuffle.
Figure~\ref{fig:hull-dnc} illustrates one merge step.

\section{Inefficiencies of the Per-Phase CUDA Decomposition}
\label{app:warp-inefficiencies}

\begin{figure}[!htbp]
  \centering
  \includegraphics[width=\linewidth]{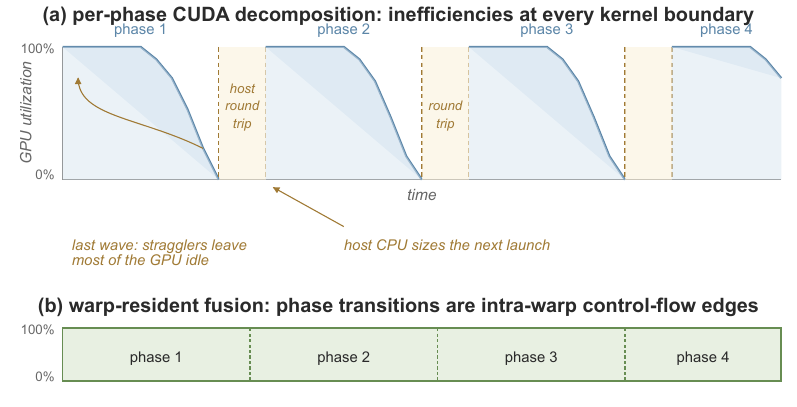}
    \vspace{-2em}
  \caption{Two inefficiencies of the per-phase CUDA decomposition.
    \textbf{Top:} a phase's work tapers off toward its boundary;
    only a few warps remain active in the last wave while the rest
    of the GPU is idle, and the next kernel cannot launch until the
    final warp retires.  \textbf{Bottom:} the next kernel's input
    buffer must be sized from the producing kernel's output, which
    forces the host CPU into the loop between every phase pair.
    Warp-resident fusion (Sec.\@~3) replaces both kernel boundaries
    with intra-warp control-flow edges and removes the host from
    the inner loop.}
  \label{fig:last-wave}
\end{figure}

This appendix details how a conventional CUDA implementation
decomposes the mesh--plane clipper of Sec.\@~4.2 into a sequence of
separate kernel launches, and how that mapping yields the two
kernel-boundary inefficiencies named in Sec.\@~1 and Sec.\@~3 of the
main paper: last-wave underfill, and the host round trip required to
size the next launch from a variable-sized output
(Figure~\ref{fig:last-wave}).

\paragraph{One kernel per phase, by example.}
The natural CUDA decomposition maps each line of Algorithm~1 that
processes a different set of items into its own kernel.  For the clipper, this expands to roughly seven
launches per cut.  (1)~\texttt{ClassifyVertices}: one thread per
input vertex writes $\mathrm{sign}(\pi \!\cdot\! v)$ into a global
sign array.  (2)~\texttt{CollectCrossEdges}: one thread per input
triangle filters triangles that cross the plane and appends their
unordered edge keys via an \texttt{atomicAdd} counter; the output
length depends on how many triangles the plane intersects.
(3)~\texttt{SortKeys}: a library sort (CUB radix or merge) sorts the
key array, after which (4)~\texttt{DedupAndIntersect} adds one
intersection vertex per unique key; the output length depends on
key uniqueness.  (5)~\texttt{SplitTriangles}: one thread per input
triangle emits up to three child triangles into $T^{+}$ or $T^{-}$;
the per-side counts are data-dependent.  (6)~\texttt{FindBoundary}:
a second key-sort plus a neighbor-scan flags unpaired boundary
edges of the smaller side; the boundary count depends on the cut
geometry.  (7)~\texttt{ChainAndCap}: chain the boundary edges into
closed loops and ear-clip them; loops are typically few and serial
enough that this last phase either runs on the host or under-fills
the GPU.  Each pair of adjacent kernels above is separated by a
device synchronization, a host-side read of the producing kernel's
output size, and a launch of the consuming kernel sized accordingly.

\paragraph{Last-wave underfill.}
A kernel scheduled at thread granularity finishes when its last
thread retires.  When a phase's work distribution shrinks toward
the kernel boundary---\texttt{CollectCrossEdges} keeps only
triangles that cross the plane, the prefilter pass of the hull
discards interior points, a tree-search node prunes most of its
children---the trailing fraction of warps runs to completion while
almost the whole machine sits idle.  The seven-launch split above
adds one such tail per kernel boundary; the fractions accumulate.

\paragraph{Variable-sized output.}
Each kernel's output size is known only after that kernel finishes
running, and in the conventional split the host CPU reads the size
and issues the next launch.  Each round trip costs a CUDA stream
synchronization and roughly the latency of \texttt{cudaMemcpy} on a
tiny payload, but issued once per phase boundary across hundreds of
thousands of cuts the cumulative cost is substantial.  CUDA Graphs
do not lift this dependency: each cut changes a phase's output
size, which changes the consuming kernel's launch shape and forces
the graph to be rebuilt.

\section{Warp-Centric Primitive Templates}
\label{app:warp-primitives-recipe}

This appendix illustrates the four primitive cases of the
warp-centric design recipe in Sec.\@~3 of the main paper.  Two of
the four already have dedicated figures elsewhere in the supplemental
and are cross-referenced rather than redrawn.

\begin{figure}[t]
  \centering
  \includegraphics[width=\linewidth]{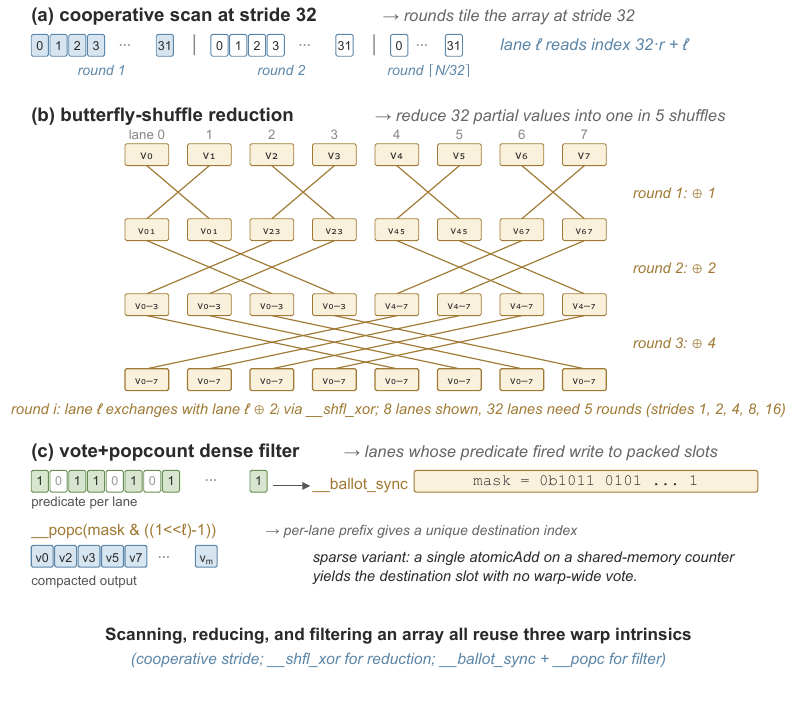}
    \vspace{-3em}
  \caption{Warp-cooperative array operations.  \textbf{Top:} stride-32
    cooperative scan over an $N$-element array; lane~$\ell$ visits
    index $32r + \ell$ on round~$r$.  \textbf{Middle:}
    butterfly-shuffle reduction combines the 32 partial values in
    $5$ shuffles via \texttt{\_\_shfl\_xor\_sync}.  \textbf{Bottom:}
    a vote+popcount dense filter compacts predicate-true elements
    into a packed output array; a sparse filter substitutes a single
    \texttt{atomicAdd} on a shared-memory counter.}
  \label{fig:warp-scan}
\end{figure}

\paragraph{(a) Scanning an array.}
The 32 lanes scan an array of $N$ elements at stride 32, so each
lane visits index $32r + \ell$ on round $r$
(Figure~\ref{fig:warp-scan}, top).  A subsequent reduction over the
32 partial values is implemented as a butterfly tree of warp
shuffles in $\log_2 32 = 5$ rounds, where on round $i$ each lane
swaps its accumulator with lane $\ell \oplus 2^{i}$ via
\texttt{\_\_shfl\_xor\_sync}~\cite{Blelloch1989Scans}
(Figure~\ref{fig:warp-scan}, middle).  A subsequent dense filter
keeps only those elements whose predicate fires:
\texttt{\_\_ballot\_sync} packs the per-lane predicate bits into a
$32$-bit mask, and the prefix popcount
$\texttt{\_\_popc}\!\bigl(\mathit{mask}\,\&\,((1\!\ll\!\ell)-1)\bigr)$
gives each surviving lane a unique destination slot in the output
(Figure~\ref{fig:warp-scan}, bottom).  When the surviving fraction
is small, a single \texttt{atomicAdd} on a shared-memory counter
yields the destination slot directly without the warp-wide vote.
The mechanics of \texttt{\_\_ballot\_sync}, \texttt{\_\_popc}, and
\texttt{\_\_shfl\_sync} are illustrated in
Figure~\ref{fig:warp-primitives}.

\paragraph{(b) Sorting, deduplication, and membership query.}
All three reduce to the warp-cooperative quick-sort of Appendix~D,
whose three-way partition is itself a vote+popcount filter
(case~(a)).  Set construction extends the sort with a single
parallel neighbor-scan that drops adjacent duplicates---an
application of the dense filter primitive on the sorted array.
A membership query against a constructed set is answered by a
binary search executed cooperatively across the 32 lanes, which
together resolve five comparison rounds before the search range
shrinks below one element.  Figure~\ref{fig:warp-primitives} of
Appendix~D illustrates the partition mechanics that carry all three
operations.

\begin{figure}[t]
  \centering
  \includegraphics[width=\linewidth]{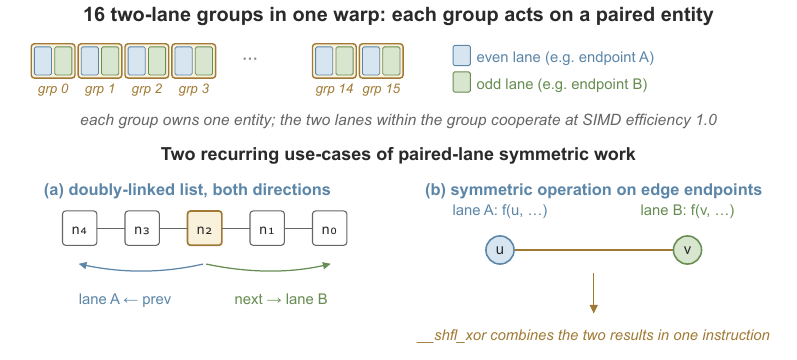}
    \vspace{-2em}
  \caption{Fine-grained symmetric work pattern.  \textbf{Top:} a
    warp partitions into 16 two-lane groups, with one lane per
    endpoint of a paired entity (e.g.\ edge endpoints or doubly
    linked list neighbors).  \textbf{Bottom:} two recurring
    use-cases---bidirectional traversal of a doubly linked list
    (left), and a symmetric operation on the two endpoints of an
    edge whose results are combined by a single warp shuffle
    (right).}
  \label{fig:warp-pair}
\end{figure}

\paragraph{(c) Fine-grained symmetric work.}
When an operation acts in parallel on two related entities---both
endpoints of an edge, both directions of a doubly linked list, both
sides of a bridge edge during convex-hull merge---we partition the
warp into 16 two-lane groups, with one lane per entity in the pair
(Figure~\ref{fig:warp-pair}).  Each group owns one independent task
and the two lanes within the group cooperate at full SIMD
efficiency: a single \texttt{\_\_shfl\_xor\_sync} exchanges register
values between the paired lanes in one instruction.  The pattern is
particularly valuable when the pair structure is the unit of work
itself, in which case 16 such units fit in one warp's
register file.  CuACD applies this template inside the convex-hull
merge of Sec.\@~4.3 (Appendix~G), where each two-lane group rotates
one bridge endpoint while the partner lane tracks the other.

\paragraph{(d) Divide and conquer.}
A $K$-way split (typically $K\!=\!16$ or $K\!=\!32$) assigns one
branch per lane, or per case-(c) lane-pair, executes the branches
in parallel, and merges the results pairwise up a tree.  CuACD
applies the pattern to the top-level split of the divide-and-conquer
hull (Sec.\@~4.3): the input point set is split $16$ ways, with one
$2$-lane group of case~(c) per chunk hulling its points in
registers, and the $16$ sub-hulls then merge pairwise up the tree.
Figure~\ref{fig:hull-dnc} of Appendix~G illustrates the
split--hull--merge sequence.  The pattern sits last in the order of
preference because the data-dependent control flow inside each
branch is not guaranteed to retain warp lockstep, eroding the SIMD
benefit unless the branches are short and structurally similar.

\clearpage

\end{document}